\documentclass[onecolumn]{aaprep} 

\usepackage{natbib}
\bibpunct{(}{)}{;}{a}{}{,} 

\usepackage{graphicx}
\usepackage{graphics}
\usepackage{subfigure}
\usepackage{txfonts}
\usepackage{longtable, lscape}
\usepackage{amscd}
\usepackage{url}

\usepackage{color}

\newcommand{\ve}[1]{\mbox{\boldmath$#1$}}
\let\vec=\ve

\def\mi{\buildrel\circ\over\i}

\def\muas{\hbox{$\mu$as}}

\begin{document}

   \title{Analysis of astrometric catalogues with vector spherical harmonics}

   \subtitle{}

   \author{F. Mignard\inst{1}\and S. Klioner\inst{2} }

   \offprints{F. Mignard, \email{francois.mignard@oca.eu}}
   \institute{Univ. Nice Sophia-Antipolis, CNRS, OCA,
              Le Mont Gros, BP 4229, 06304 Nice Cedex 4, France\\
               \email{francois.mignard@oca.eu} \and
              Lohrmann Observatory, Technische Universit\"at Dresden, 01062, Dresden, Germany\\
               \email{sergei.klioner@tu-dresden.de}
             }
   \date{Received \today / Accepted \today}

  \abstract
  {}
   {We compare stellar catalogues with position and proper motion
    components using a decomposition on a set of orthogonal vector
    spherical harmonics. We aim to show the theoretical and practical
    advantages of this technique as a result
    of invariance properties and the independence of the
    decomposition from a prior model.}
   {We describe the mathematical principles used to perform the
     spectral decomposition, evaluate the level of significance of
     the multipolar components, and examine the transformation properties under space rotation.}
   {The principles are illustrated with a characterisation of  systematic
    effects in the FK5 catalogue compared to Hipparcos and with an
    application to extraction of the rotation and dipole acceleration in
    the astrometric solution of QSOs expected from Gaia.}
   {}

   \titlerunning{Astrometric catalogue analysis}
   \maketitle
%

\section{Introduction} \label{sect:intro}

The differences between the positions of a set of common sources in two
astrometric catalogues are conveniently described mathematically by a
vector field on a sphere. Each vector materialises the difference
between the two unit vectors giving the direction of the common
sources in each catalogue. This feature extends to the differences
between the proper motions of each source which also generates a
spherical vector field. Typically when connecting two positional
catalogues with $n$ common sources, one has the coordinates
$\alpha_i^{(1)}, \delta_i^{(1)}$ for the $i$th source in the first
catalogue and $\alpha_i^{(2)}, \delta_i^{(2)}$ for the same source in
the second catalogue. Provided the two catalogues are close to each
other, the differences can be mapped as a vector field with components
in the local tangent plane given by
\begin{equation}\label{vfintro}
    \vec{V} =  [\Delta\alpha\cos\delta = (\alpha^{(2)} -\alpha^{(1)})\cos\delta^{(1)} ,\ \Delta\delta =\delta^{(2)} -\delta^{(1)}]
\end{equation}
\noindent
for each common source between the two catalogues. We wish to analyse
these fields in order to summarise their largest overall or local
features by means of a small set of base functions, which is much
smaller in any case than the number of sources.

Several papers in the astronomical literature have applied this
overall idea with either scalar or vectorial functions. As far as we
have been able to trace the relevant publications regarding the use of
vector spherical harmonics in this context, this has been initiated by
\citet{Mignard90} and published in a proceedings paper that is not
widely accessible. One of the motivations of this paper is therefore
to update and expand on this earlier publication and to provide more
technical details on the methodology.

The general idea of the decomposition has its root in the classical
expansion of a scalar function defined on the unit sphere with a set
of mutually orthogonal spherical harmonics functions. The most common
case in astronomy and geodesy is the expansion of the gravitational
potential of celestial bodies, in particular that of the Earth. The
generalisation of such expansions to vectors, tensors, or even
arbitrary fields was introduced in mathematics and theoretical physics
decades ago \citep{GelfandMilnosShapiro1963}.  For example, these
generalisations are widely used in gravitational physics
\citep{Thorne1980,Suen1986,blanchet86}. Restricting it to the case of
vectors fields, which is the primary purpose of this paper, we look
for a set of base functions that would allow representing any vector field on
the unit sphere as an infinite sum of fields, so that the angular
resolution would increase with the degree of the representation
(smaller details being described by the higher harmonics).  One would
obviously like for the lowest degrees to represent the most
conspicuous features seen in the field, such as a rotation about any
axis or systematic distortion on a large scale. It is important to
stress at this point that expanding a vector field on this basis
function is not at all the same as expanding the two scalar fields
formed by the components of the vector field: the latter would depend
very much on the coordinate system used, while a direct representation
on a vectorial basis function reveals the true geometric properties of
the field, regardless of the coordinate system, in the same way as the
vector field can be plotted on the sphere without reference to a
particular frame without a graticule drawn in the background.

Several related methods using spherical functions to model the errors
in astrometric catalogues, the angular distribution of proper
motions, or more simply as a means of isolating a rotation have been
published and sometimes with powerful
algorithms. \citet{brosche66,brosche70} was probably the first to
suggest the use of orthogonal functions to characterise the errors in
all-sky astrometric catalogue, and his method was later improved by
\citet{schwan77} to allow for a magnitude dependence. But in both
cases the analysis was carried out separately for each component
$\Delta\alpha\cos\delta, \Delta\delta$ of the error, by expanding two
scalar functions. Therefore, this was not strictly an analysis of a
vector field, but that of its components on a particular reference
frame. The results therefore did not describe the intrinsic geometric
properties of the field, which should reveal properties not connected
to a particular frame (see also Appendix \ref{annex_2}). Similarly, a
powerful method for exhibiting primarily the rotation has been developed
in a series of papers of \citet[and references therein]{Vityazev2010}.
The vector spherical harmonics (hereafter abridged in VSH) have been
in particular used in several analyses of systematic effects in VLBI
catalogues or in comparison of reference frames, such as in
\citet{arias2000}, \citet{titov2009}, or \citet{gwinn97}, the galactic
velocity field \citep{MakarovEtAl2007}, the analysis of zonal errors
in space astrometry \citep{MakarovEtAl2012}, but none of these
papers deals with the fundamental principles, the very valuable
invariance properties and the relevant numerical methods. Within the
Gaia preparatory activities the VSH are used also to compare sphere
solutions and in the search of systematic differences at different
scales by \citet{bucciarelli2011} and more is planned in the future to
characterise and evaluate the global astrometric solution.

In this paper we provide the necessary theoretical background to
introduce the VSHs and the practical formulas needed to explicitly compute
the expansion of a vector field. The mathematical
principles are given in Section~\ref{sect:math} with a few illustrations
to show the harmonics of lower degrees. In the next section,
Section~\ref{wigner}, we emphasise the transformation of the VSHs and
that of the expansion of a vector field under a rotation of the
reference frame, with applications to the most common astronomical
frames. Then in Section~\ref{sect:global} we discuss the physical
interpretation of the harmonics of first degree as the way of
representing the global effect shown by a vector field, such as the axial
rotation and its orthogonal counterpart, for which we have coined the
term \textsl{glide}, and show its relationship to the dipole axial
acceleration resulting from the Galactic aberration. The practical
implementation is taken up in Section~\ref{implementation}, where we also
discuss the statistical testing of the power found in each
harmonic. The results of particular applications to the FK5 Catalogue
and to the simulated QSO catalogue expected from Gaia are respectively
given in Sects.~ \ref{sect:fk5} and \ref{sect:Gaia-results}. Appendix
\ref{annex_1} provides the explicit expressions of the VSH up to
degree $l=4$, while Appendix \ref{annex_numerical} deals with the
practical formulas needed for numerical evaluation of the VSH, and
Appendix \ref{annex_2} focuses on the relationship between expansions
of the components of a vector field on the scalar spherical harmonics
and the expansion of the same field on the VSHs.

\section{Mathematical principles}
 \label{sect:math}

  \begin{figure}[htb]
   \centering \vspace{0.0cm}
 \includegraphics[width=0.6\hsize]{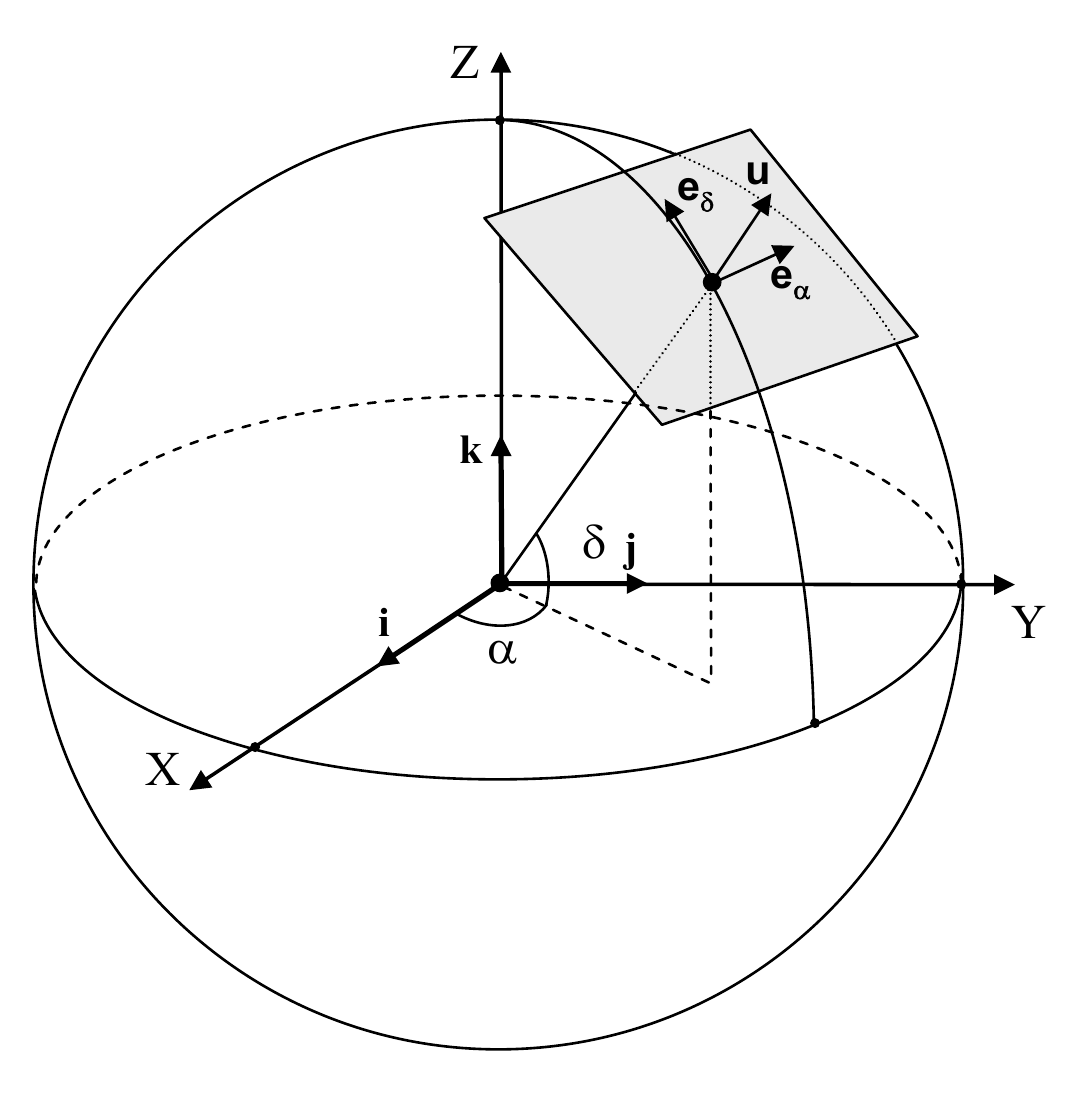}
      \caption{Local frame associated to the spherical coordinates $\alpha, \delta$, with the unit vectors along the longitude and latitude lines.}
       \label{fig:local_frame}
   \end{figure}

\subsection{Definitions}

In this section we give the main definitions needed to use the
VSHs and the theoretical background from which practical numerical
methods are established and discussed later in the paper. The VSHs
form an orthogonal set of basis functions for a vector field on a
sphere and split into two categories referred to as the toroidal
$\ve{T}_{lm}$ and spheroidal $\ve{S}_{lm}$ functions (in the physical
literature the former are also called ``magnetic'' or ``stream'',
and the latter ``poloidal'', ``potential'', or ``electric''):
\begin{eqnarray}
\ve{S}_{lm}&=&\frac{1}{\sqrt{l(l+1)}}\,r\,\ve{\nabla} Y_{lm}=\ve{u}\times\ve{T}_{lm}\,,
\\
\label{T=-uxS}
\ve{T}_{lm}&=&-\,\ve{u}\times\ve{S}_{lm}\,,
\end{eqnarray}
\noindent
where $\ve{u}=\ve{r}/r$, $r=|\ve{r}|$, $\ve{r}$ is the
radius-vector of the point, and $\ve{\nabla}$ denotes the gradient operator.
Clearly, for points on the surface of a unit sphere $r=1$. It is also obvious
that
at each point on the sphere and for each
$l$ and $m$ one has
$\ve{u}\cdot\ve{S}_{lm}=0$, $\ve{u}\cdot\ve{T}_{lm}=0$, and
$\ve{T}_{lm}\cdot\ve{S}_{lm}=0$.
Taking into account that
(see Fig. \ref{fig:local_frame})
\begin{eqnarray}
\label{ni-alpha-delta}
\ve{u}&=&\pmatrix{\cos\alpha\,\cos\delta\cr
\sin\alpha\,\cos\delta\cr
\sin\delta}\,,
\\
\label{e_alpha}
\ve{e}_\alpha&=&{1\over\cos\delta}\,{\partial\over\partial\alpha}\ve{u}
=\pmatrix{-\sin\alpha\cr \phantom{-}\cos\alpha\cr 0}\,,
\\
\label{e_delta}
\ve{e}_\delta&=&\ve{u}\times\ve{e}_\alpha={\partial\over\partial\delta}\,\ve{u}
=\pmatrix{-\cos\alpha\,\sin\delta\,\cr -\sin\alpha\,\sin\delta\cr \cos\delta},
\end{eqnarray}
\noindent
(so that $|\ve{u}|=1$, $|\ve{e}_\alpha|=1$, $|\ve{e}_\delta|=1$),
one gets explicit formulas
 \begin{equation}\label{Tlm}
    \ve{T}_{lm}(\alpha,\delta) = \frac{1}{\sqrt{l(l+1)}} \left[\frac{\partial Y_{lm}}{\partial\delta}\ve{e}_\alpha - \frac{1}{\cos\delta}\frac{\partial Y_{lm}}{\partial\alpha}\ve{e}_\delta \right]
 \end{equation}
\noindent
for the toroidal functions, and
 \begin{equation}\label{Slm}
    \ve{S}_{lm}(\alpha,\delta) = \frac{1}{\sqrt{l(l+1)}} \left[\frac{1}{\cos\delta}\frac{\partial Y_{lm}}{\partial\alpha}\ve{e}_\alpha +\frac{\partial Y_{lm}}{\partial\delta}\ve{e}_\delta  \right]
 \end{equation}
\noindent
for the spheroidal functions. The $Y_{lm}$ are the standard spherical
functions defined here with the following sign convention
\begin{equation}\label{ylm}
Y_{lm}(\alpha,\delta) = (-1)^m \, \sqrt{\frac{2l+1}{4\pi}\,\frac{(l-m)!}{(l+m)!}}\, P_{lm}(\sin\delta)\,e^{\mi m\alpha}
\end{equation}
\noindent
for $m \geq 0$ and
\begin{equation}\label{ylminusm}
Y_{l,-m}(\alpha,\delta) = (-1)^m \, Y_{lm}^{\ast}(\alpha,\delta)
\end{equation}
\noindent
for $m < 0$, where superscript `$\ast$' denotes complex conjugation.
The associated Legendre functions are defined as
\begin{equation}\label{Plm}
    P_{lm}(x) = (1-x^2)^{m/2}\,\frac{d^m P_l(x)}{dx^m}.
\end{equation}
\noindent
Equations (\ref{ylm})--(\ref{ylminusm}) agree with the well-known formula
for the associated Legendre functions
\begin{equation}\label{Pl-m}
    P_{l,-m}(x) = (-1)^m\,{(l-m)!\over (l+m)!}\,P_{lm}(x)\,.
\end{equation}
\noindent
Different sign conventions appear in the literature, in particular
regarding the place of the $(-1)^m$, which is sometimes used in the
definition of the associated Legendre functions instead
\citep[see, e.g., Chapter 8 of][]{AbramowitzStegun1972}.
However, these sign
differences do not influence the vector spherical functions
themselves. Also one sometimes considers $\mi\,\ve{T}_{lm}$
instead of $\ve{T}_{lm}$ as the toroidal vector spherical functions.

From (\ref{ylminusm}) and (\ref{Tlm})--(\ref{Slm}) one immediately has
\begin{eqnarray}\label{conjugated}
\ve{T}_{l,-m}(\alpha,\delta) &=& (-1)^m \, \ve{T}_{lm}^{\ast}(\alpha,\delta)\,,
\\
\ve{S}_{l,-m}(\alpha,\delta) &=& (-1)^m \, \ve{S}_{lm}^{\ast}(\alpha,\delta)\,.
\end{eqnarray}

\subsection{Orthogonality properties}\label{subsec:properties}

The spherical functions $Y_{lm}$ form an orthonormal sequence of functions
on the surface of a sphere since
\begin{equation}\label{YdotY}
    \int_\Omega\, Y_{lm}Y_{l'm'}^{\ast}\,d\Omega= \delta^{ll'}\delta^{mm'}\,,
\end{equation}
which is also complete in the Hilbert space $\mathcal{S}$ of the
square-integrable functions on a sphere (making it an
orthonormal basis of $\mathcal{S}$  in the $L^2$ norm). As in
Fourier expansions, the completeness property is hard to establish and
is related to the fact that spherical harmonics are eigenfunctions of
a special kind of differential equations  \citep[see, for
  example,][]{MorseFeshbach1954}.  This is accepted and not
further discussed in this paper directed towards astronomical
applications. Here and below one has
$d\Omega=\cos\delta\,d\delta\,d\alpha$, and the integration is
taken over the surface of the unit sphere: $0\le\alpha\le2\pi$,
$-\pi/2\le\delta\le\pi/2$, and $\delta^{ij}$ is the Kronecker symbol:
$\delta^{ij}=1$ for $i=j$ and $\delta^{ij}=0$, otherwise.

Thanks to the completeness property a (square-integrable)
complex-valued scalar function defined on a unit sphere $f(\alpha,\delta)$
can be uniquely projected on the $Y_{lm}$:
\begin{equation}\label{f-Ylm}
    f(\alpha,\delta)=\sum_{l=0}^\infty\sum_{m=-l}^l f_{lm}\,Y_{lm}\,,
\end{equation}
\noindent
where the Fourier coefficients $f_{lm}$ age given by
\begin{equation}\label{fourier}
f_{lm} = \int_\Omega\,f\,Y^{\ast}_{lm}\,d\Omega\,.
\end{equation}
\noindent
The equality in (\ref{f-Ylm}) strictly means that the
right-hand side series converge (not necessarily pointwise
at discontinuity points or singularities but at
least with the $L^2$ norm) to some function $f$, and given the
completeness and the orthonormal basis, the coefficients $f_{lm}$ are
related to $f$ by (\ref{fourier}). A truncated form of (\ref{f-Ylm})
with $l \leq l_\mathrm{max} < \infty$ is an approximate expansion for
which the equality in (\ref{f-Ylm}) does not strictly hold (in
general).

Similarly, the VSH form a complete set of orthonormal vector functions
on the surface of a sphere (with the inner product of the $L^2$
space):
\begin{equation}\label{TdotT}
    \int_\Omega\, \ve{T}_{lm}\cdot \ve{T}^{\ast}_{l'm'}\,d\Omega= \delta^{ll'}\delta^{mm'}\,,
\end{equation}
\begin{equation}\label{SdotS}
    \int_\Omega\, \ve{S}_{lm}\cdot \ve{S}^{\ast}_{l'm'}\,d\Omega= \delta^{ll'}\delta^{mm'}\,,
\end{equation}
\begin{equation}\label{SdotT}
    \int_\Omega\, \ve{S}_{lm}\cdot \ve{T}^{\ast}_{l'm'}\,d\Omega= 0\,.
\end{equation}
\noindent
As we noted after (\ref{T=-uxS}) above, one also has the orthogonality
between two vectors in the usual Euclidean space,
\begin{equation}\label{STeuclid}
    \ve{S}_{lm}\cdot \ve{T}_{lm} = 0
\end{equation}
\noindent
which holds for any point on the sphere.

Any (square-integrable) complex-valued vector field $\ve{V}(\alpha,\delta)$
defined on the surface of a sphere and orthogonal to $\ve{u}$ (radial direction),
\begin{equation}\label{field}\label{Vexpandcomp}
    \vec{V}(\alpha,\delta) = V^\alpha(\alpha,\delta)\,\vec{e}_\alpha
+  V^\delta(\alpha,\delta)\,\vec{e}_\delta\,,
\end{equation}
\noindent
can be expanded in a unique linear combination of the VSH,
\begin{equation}\label{Vexpand}
    \ve{V}(\alpha,\delta) = \sum_{l=1}^\infty\,\sum_{m=-l}^{l}\,
\bigl(t_{lm} \ve{T}_{lm} + s_{lm} \ve{S}_{lm}\bigr)\,,
\end{equation}
\noindent
where the coefficients $t_{lm}$ and $s_{lm}$ can again be computed by
projecting the field on the base functions with
\begin{eqnarray}\label{coef}
t_{lm} &=&  \int_\Omega\,  \ve{V}\cdot \ve{T}^{\ast}_{lm}\,d\Omega\,,\\[3pt]
s_{lm} &=&  \int_\Omega\,  \ve{V}\cdot \ve{S}^{\ast}_{lm}\,d\Omega\,.
\end{eqnarray}

\subsection{Numerical computation of the vector spherical harmonics}

Analytical expressions of the VSH are useful for the lower orders to
better understand what fields are represented with large-scale
features and also to experiment \textsl{by hand} on simple fields in
an analytical form. This helps develop insight into their properties
and behaviour, which pays off at higher orders when one must rely only
on numerical approaches. This is also a good way to test a computer
implementation by comparing the numerical output to the expected
results derived from the analytical expressions. Analytical
expressions for the vector spherical harmonics of orders $l\le4$ are
given explicitly as a function of $\alpha$ and $\delta$ in Appendix
\ref{annex_1}.

To go to higher degrees, one needs to resort to numerical methods. To
numerically compute the two components of
$\ve{T}_{lm}(\alpha,\delta)$ and $\ve{S}_{lm}(\alpha,\delta)$
given by (\ref{Tlm})--(\ref{Slm}), one first needs to have a procedure
for the scalar spherical harmonics. Given the form of the
$Y_{lm}(\alpha,\delta)$ in (\ref{ylm}), the derivative with respect to
$\alpha$ is trivial, and for $\delta$ only, the derivative of the
Legendre associated functions needs special care. There are several
recurrence relations allowing  the Legendre functions to be computed, but
not all of them are stable for high degrees. Other difficulties
appear around the poles when $\delta \pm \pi/2$ where care must be
exercised to avoid singularities. Non-singular expressions and
numerically stable algorithms for computing the VSH components are
available in the literature, and the expressions we have implemented
and tested are detailed in Appendix \ref{annex_numerical}.

\subsection{Expansions of real functions}

Describing real-valued vector fields with complex-valued VSHs
leads to unnecessary redundancy in the number of basis functions
and fitting coefficients.
In general, the coefficients in (\ref{Vexpand}) are complex even for a
real vector field. But for a real field $\ve{V}(\alpha,\delta)$, the
expansion must be real. Given the conjugation properties of the VSH,
it is clear that for a real-values function one gets
\begin{eqnarray}\label{reim1}
   t_{lm} &=& (-1)^m\,t^{\ast}_{l,-m}\,, \\
\label{reim1-s}
   s_{lm} &=& (-1)^m\,s^{\ast}_{l,-m}\,.
\end{eqnarray}
\noindent
Then from the decomposition of the coefficients into their real and
imaginary parts as
 \begin{eqnarray}\label{reim2}
   t_{lm} &=& t^{\Re}_{lm} + \mi\,  t^{\Im}_{lm}\,, \\
   s_{lm} &=& s^{\Re}_{lm} + \mi\,  s^{\Im}_{lm}\,,
\end{eqnarray}
\noindent
one gets at the end by rearranging the summation on $m \geq 0$
\begin{eqnarray}\label{Vexpandreal}
    \ve{V}(\alpha,\delta) &=&  \sum_{l=1}^\infty\,\Biggl(
t_{l0} \ve{T}_{l0} + s_{l0} \ve{S}_{l0}
\nonumber\\
&&
\phantom{\sum_{l=1}^\infty\,\Biggl(}
+ 2\sum_{m=1}^{l}\, \left(t^{\Re}_{lm} \ve{T}^{\Re}_{lm} -  t^{\Im}_{lm} \ve{T}^{\Im}_{lm}
+s^{\Re}_{lm} \ve{S}^{\Re}_{lm} -  s^{\Im}_{lm} \ve{S}^{\Im}_{lm}
\right)\Biggr)\,,
\end{eqnarray}
\noindent
which is real. It is obvious that $t^{\Im}_{l0}=s^{\Im}_{l0}=0$ and,
therefore, $t_{l0}=t^{\Re}_{l0}$ and $s_{l0}=s^{\Re}_{l0}$.

It is useful to note that the use of real and
imaginary parts of the VSHs as in (\ref{Vexpandreal}) is analogous
to the use of ``$\sin$'' and ``$\cos$'' spherical functions
suggested e.g. by \citet{brosche66}. However, we prefer to
retain notations here that are directly related to the complex-values VSHs
and the corresponding expansion coefficients.

 Equation (\ref{Vexpandreal}) provides the basic model (with only real
numbers) to compute the coefficients by a least squares fitting of a
field given for a finite number of points, which are not necessarily
regularly distributed. In this case the discretised form of the integrals
in (\ref{TdotT}), (\ref{SdotS}), and (\ref{SdotT}) is never exactly $0$
or $1$, and the system of VSH on this set of points is not fully
orthonormal. Then one cannot compute the coefficients by a direct
projection. However, one can solve the linear model
(\ref{Vexpandreal}) on a finite set of coefficients $t_{lm}$ and
$s_{lm}$. Provided the errors are given by a Gaussian noise, the
solution should produce unbiased estimates of the true
coefficients. This is discussed further in
Section~\ref{implementation}.

\subsection{Relation between the expansion in vector and scalar spherical harmonics}
Once we have fitted a vector field on the model (\ref{Vexpandreal}), we
have the expansion in the form (\ref{Vexpand}) with $l \leq
l_\mathrm{max}$.  The components $V^\alpha$ and $V^\delta$ of the
vector field $\ve{V}$ in the local basis $\ve{e}_\alpha$ and
$\ve{e}_\delta$ as in (\ref{field}) are expressed with the same set of
coefficients $t_{lm}$ and $s_{lm}$ on their respective components of
the VSH. But since the $V^\alpha$ and $V^\delta$ are scalar functions
on a sphere, they could also have been expanded independently in terms
of scalar spherical functions $Y_{lm}$ as
$V^\alpha=\sum_{l=0}^\infty\sum_{m=-l}^l V^\alpha_{lm}\,Y_{lm}$ and
$V^\delta=\sum_{l=0}^\infty\sum_{m=-l}^l V^\delta_{lm}\,Y_{lm}$,
providing another representation, which has been used as mentioned in
the Introduction to analyse stellar catalogues in the same
spirit as with the VSHs. It is therefore important to relate the two
different sets of coefficients for the purpose of comparing and
discussing the major difference between the two approaches (see
Section \ref{wigner}). This issue is not central to this paper so the
mathematical details are deferred to in Appendix \ref{annex_2}. One must,
however, note that the relations between coefficients $V^\alpha_{lm}$
and $V^\delta_{lm}$, on the one hand, and $t_{lm}$ and $s_{lm}$, on the
other, are rather complicated and involve infinite linear
combinations. It means, for example, that the information contained in a
single coefficient $t_{lm}$ is spread over infinite number of
coefficients $V^\alpha_{lm}$ and $V^\delta_{lm}$ \citep[see, e.g.][for
a particular case $l=1$]{Vityazev2010}.

\begin{figure*}[htb]
\begin{center}
\begin{tabular}{cc}
\subfigure[$S_{1,1}$]{
\includegraphics[width=8cm]{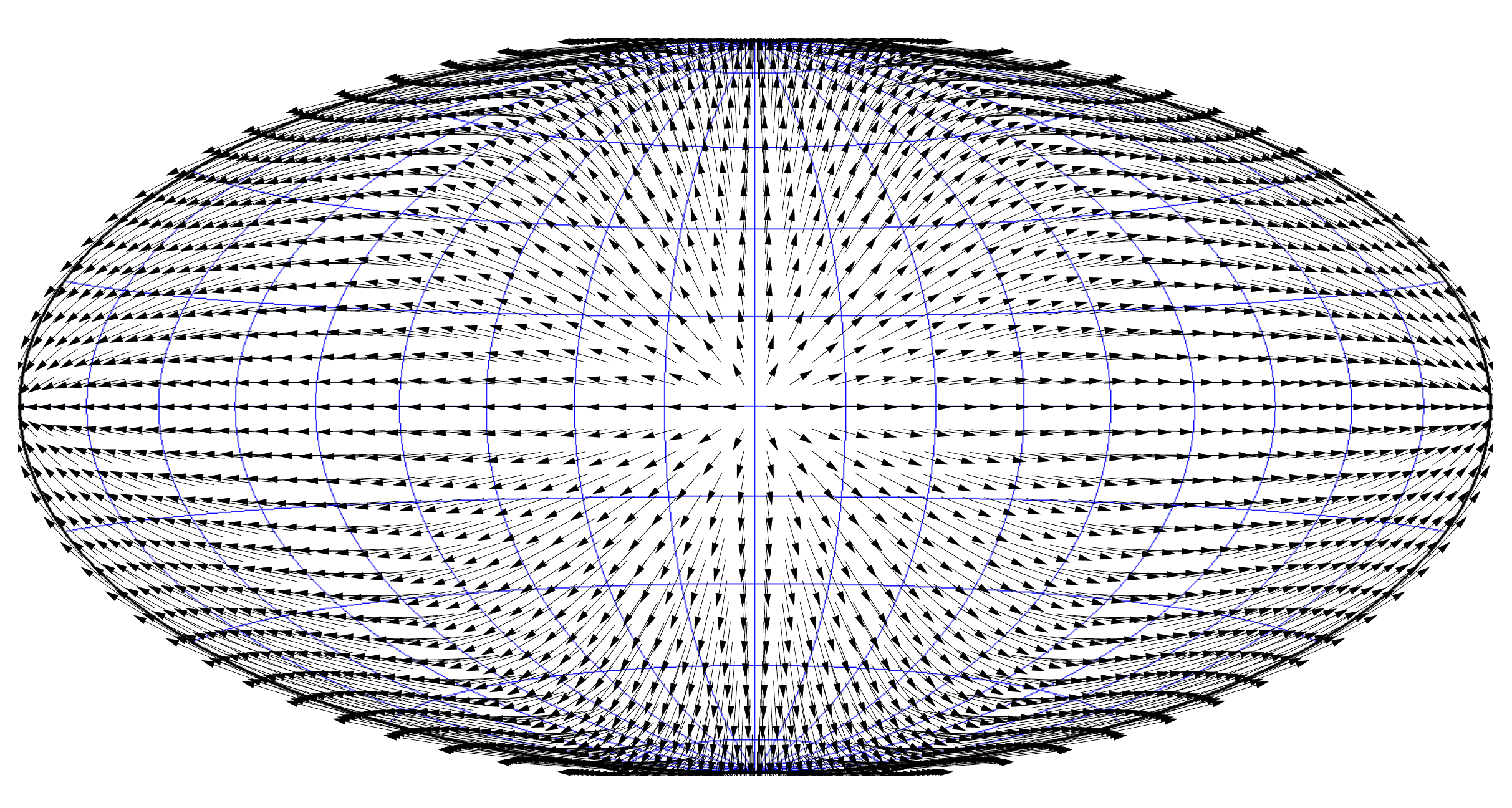}
}
&
\subfigure[$S_{2,1}$]{
 \includegraphics[width=8cm]{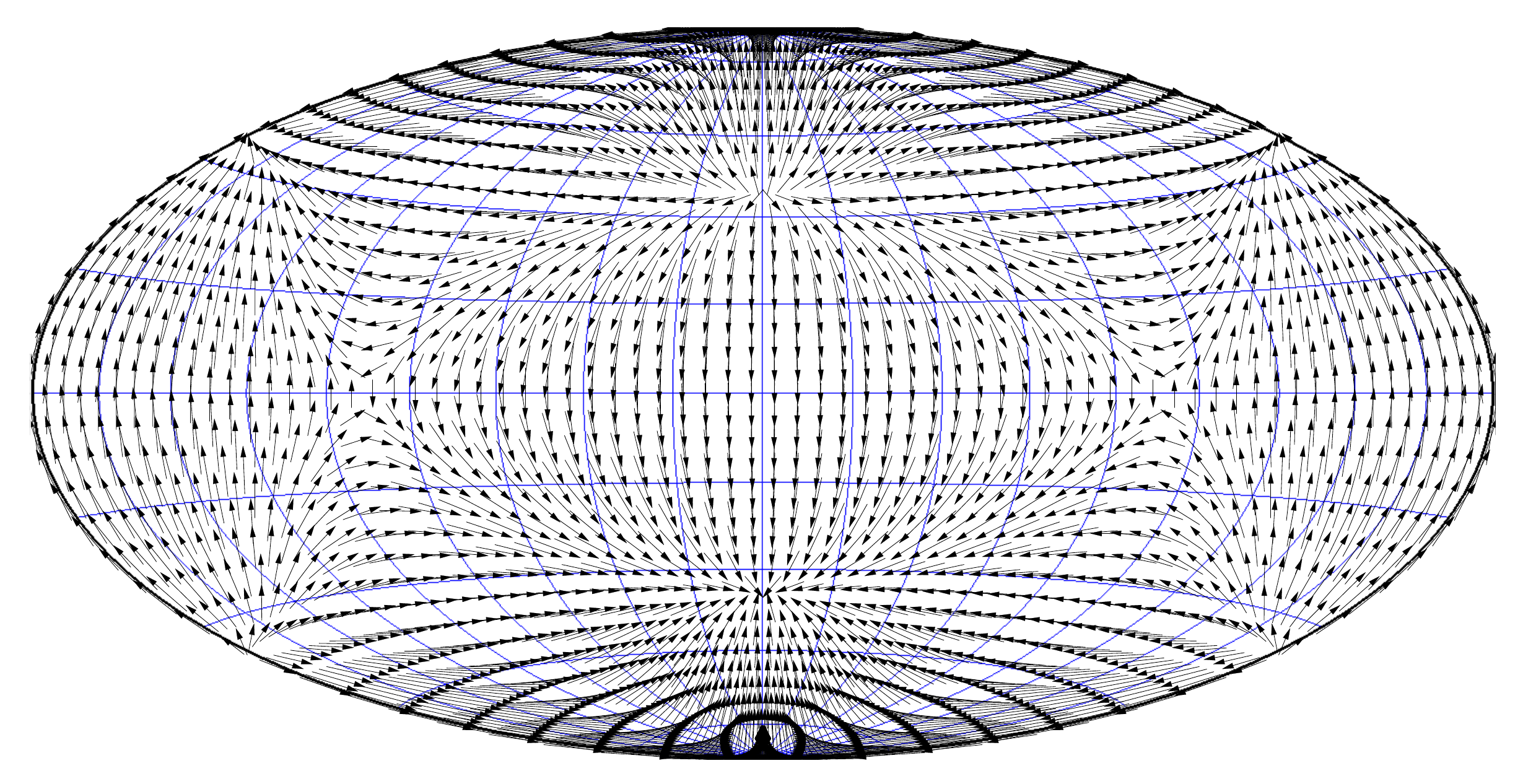}
}
\\
\subfigure[$S_{5,3}$]{
 \includegraphics[width=8cm]{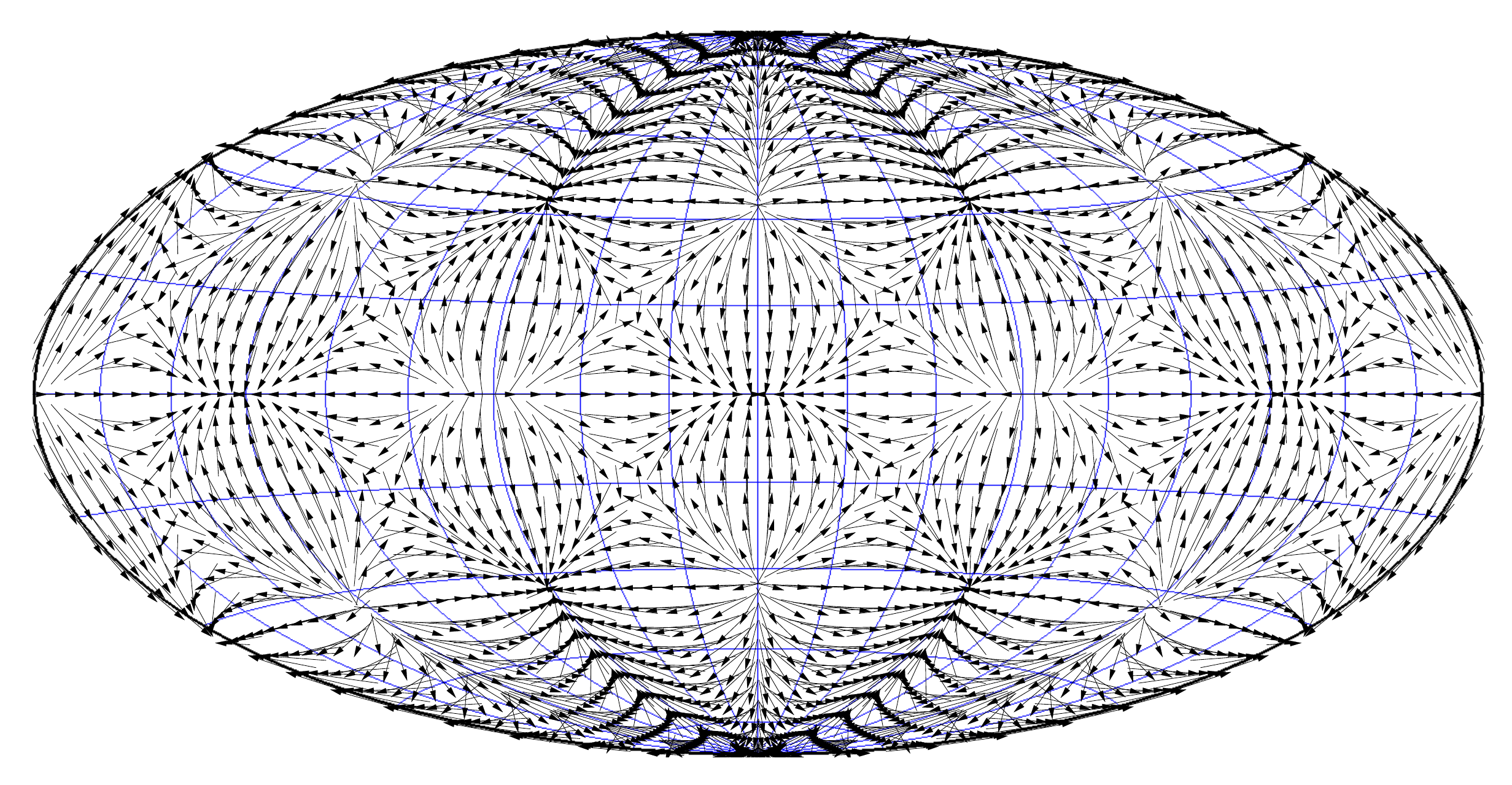}
}
&
\subfigure[$S_{10,5}$]{
 \includegraphics[width=8cm]{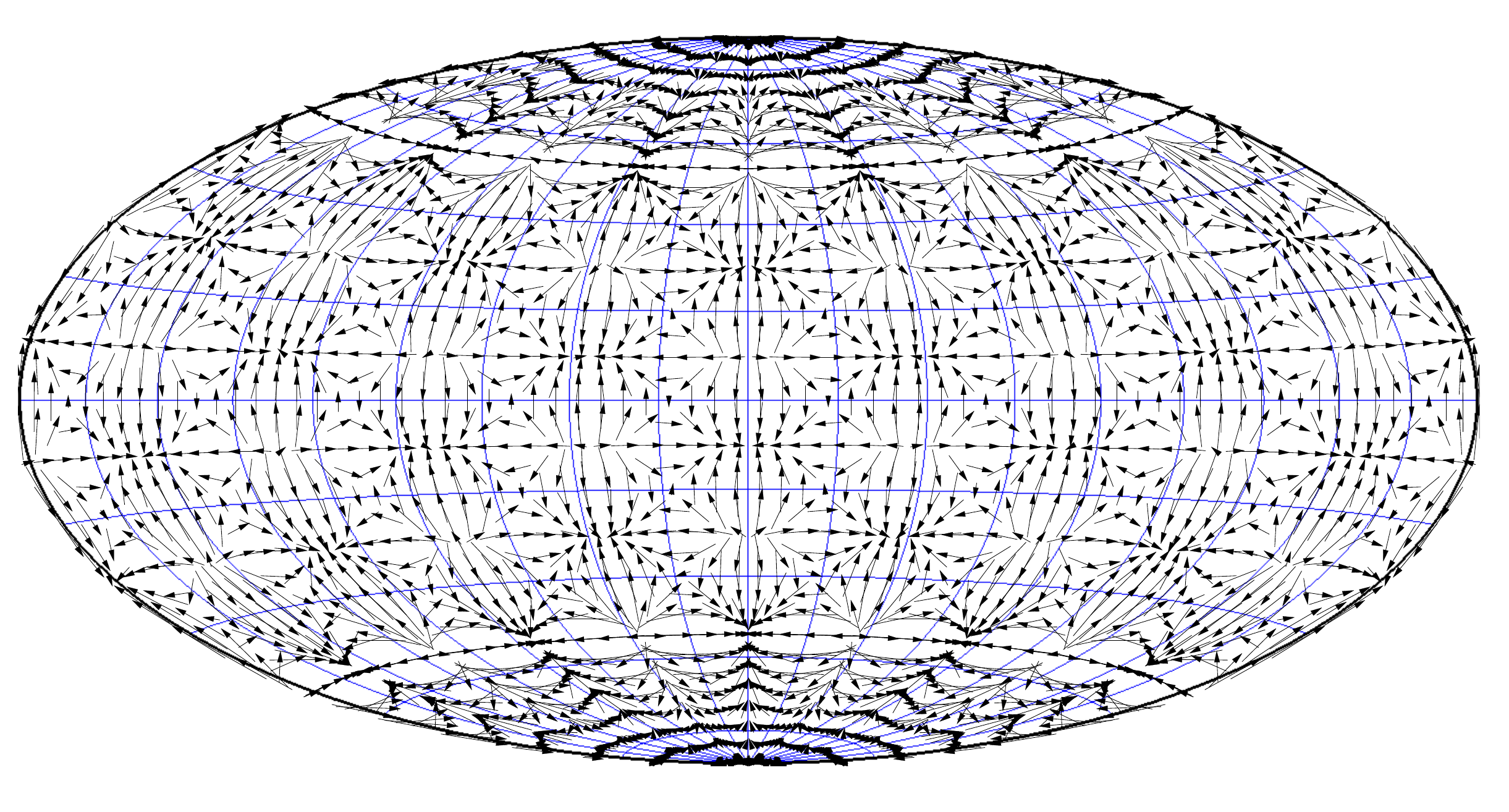}
}
\end{tabular}
\end{center}
 \caption{Examples of spheroidal harmonics vector fields:
$S_{1,1}$, $S_{2,1}$, $S_{5,3}$, $S_{10,5}$, where one
   sees the change in the angular resolution with higher degrees. The
   $S_{1,1}$ is a glide flow in the direction of the $y$-axis.}
       \label{fig:spheroidal_harm}
   \end{figure*}

 \begin{figure*}[htb]
\begin{center}
\begin{tabular}{cc}
\subfigure[$T_{1,1}$]{
 \includegraphics[width=8cm]{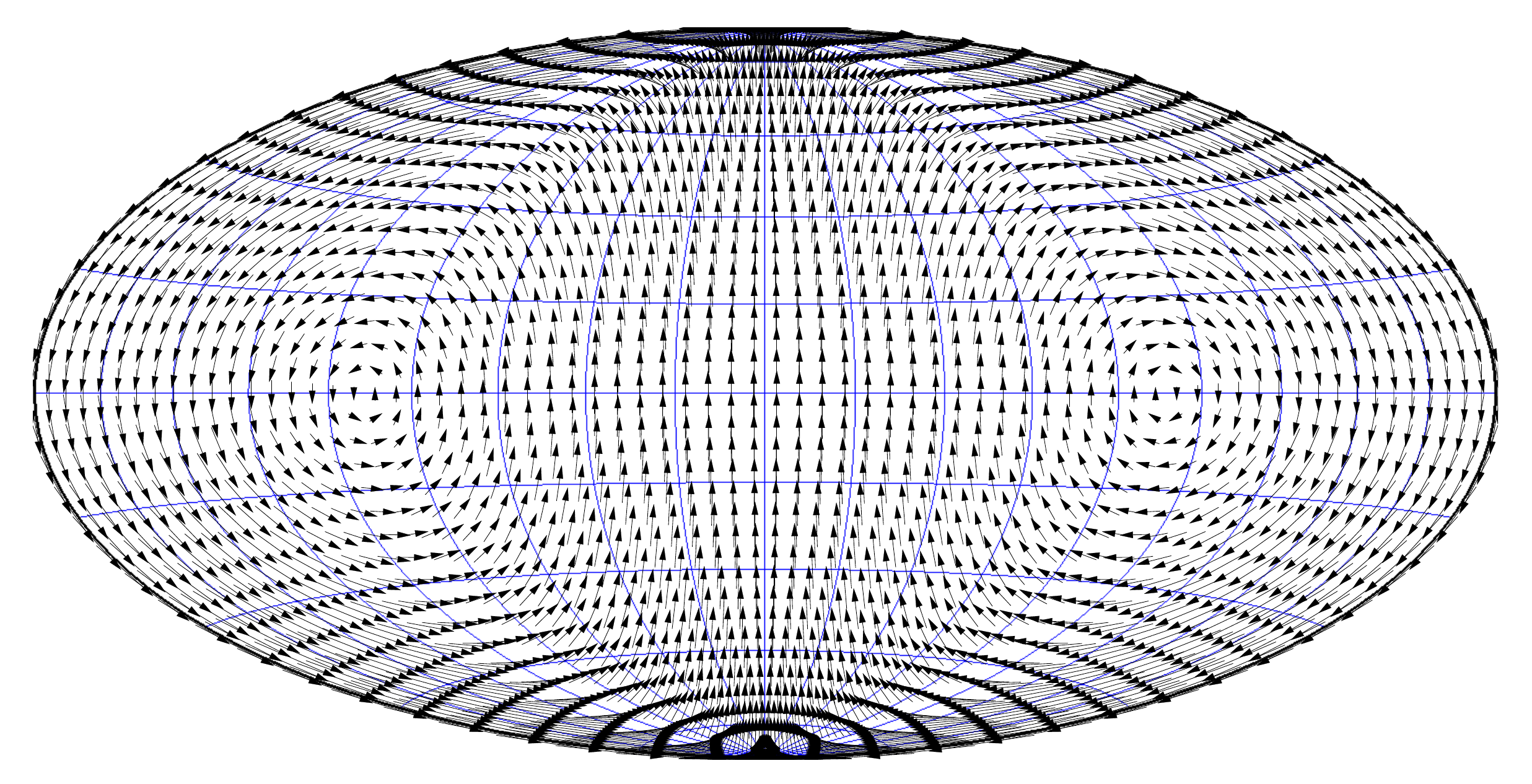}
}
&
\subfigure[$T_{2,1}$]{
 \includegraphics[width=8cm]{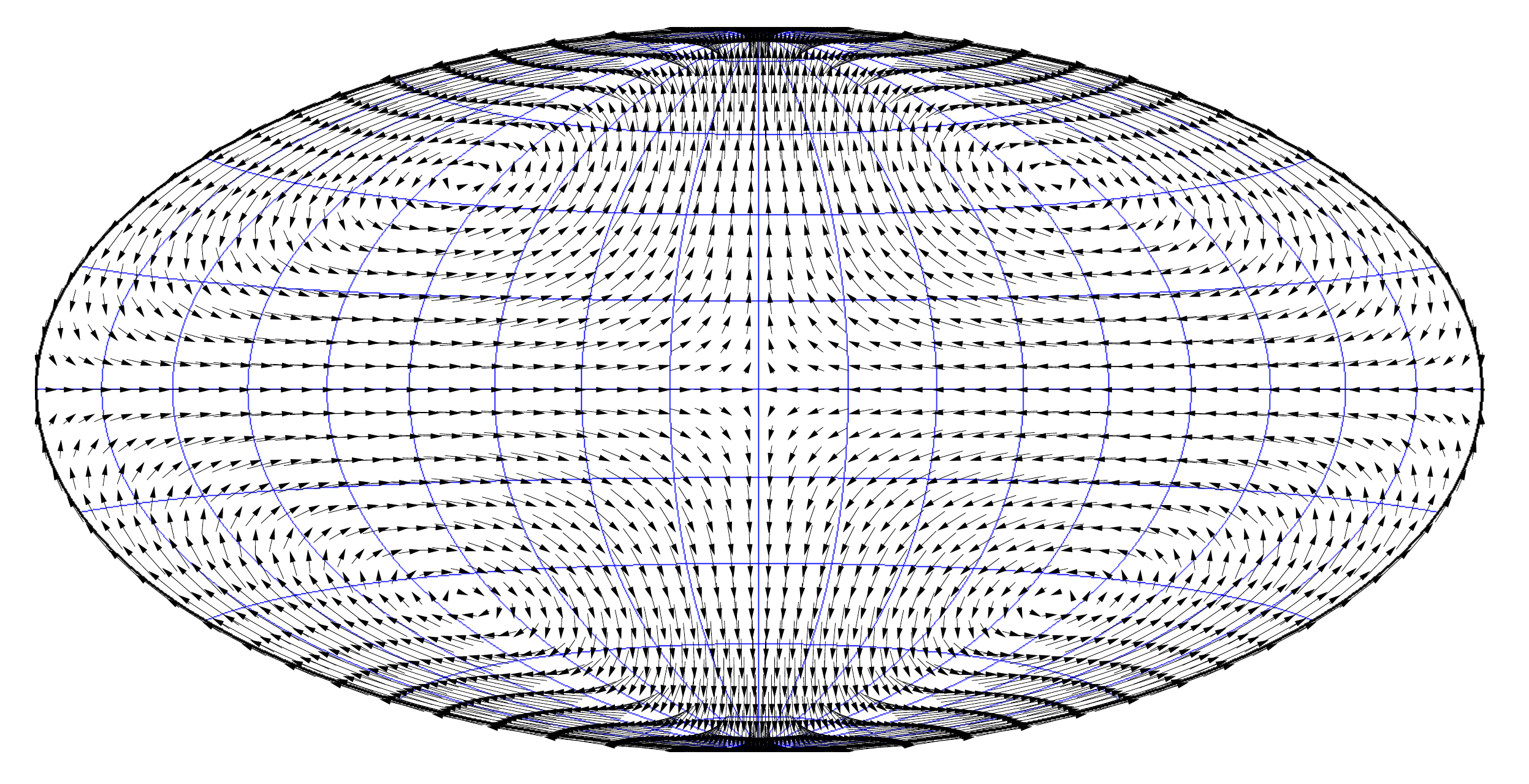}
}
\\
\subfigure[$T_{5,3}$]{
 \includegraphics[width=8cm]{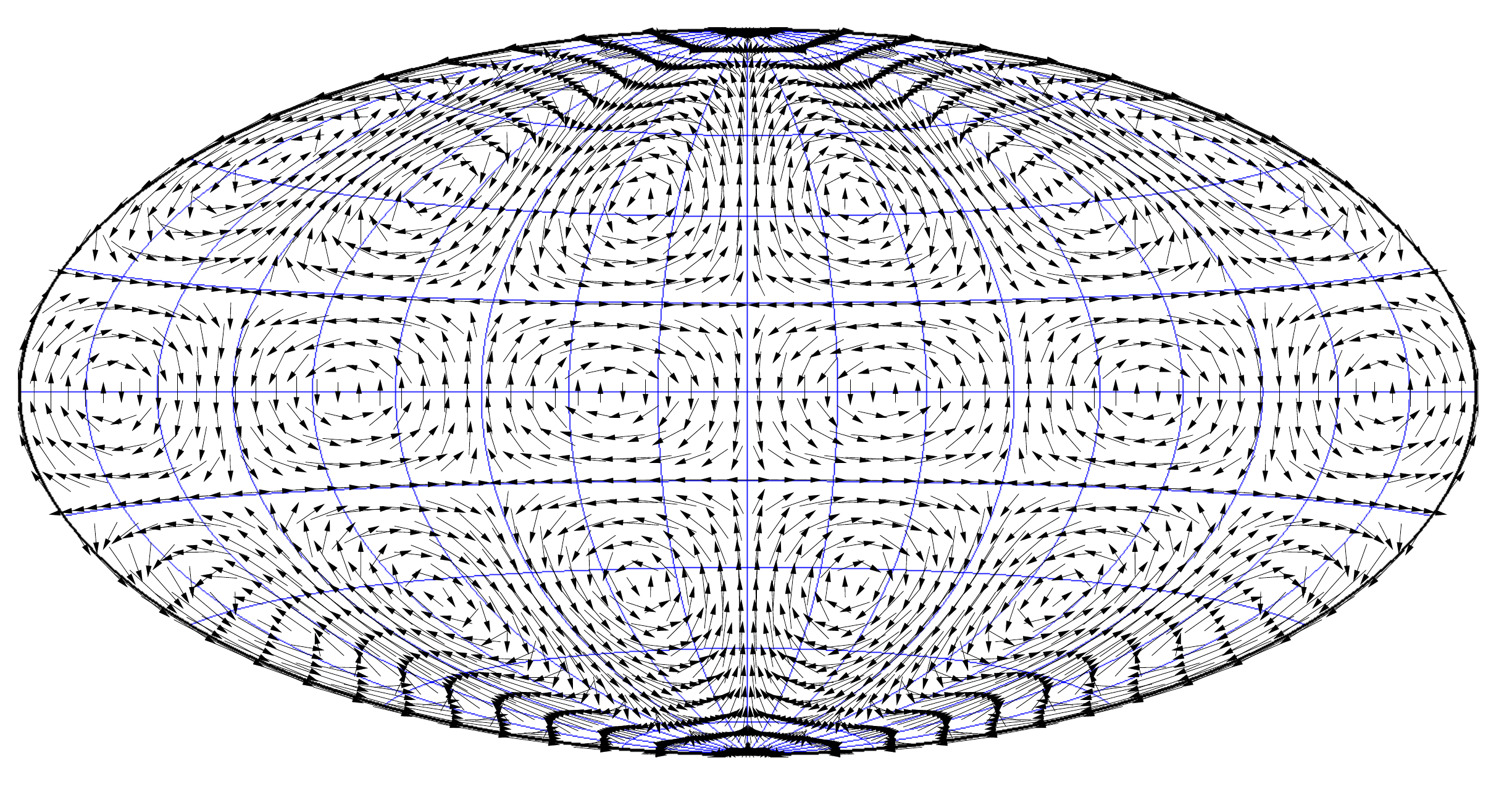}
}
&
\subfigure[$T_{10,5}$]{
 \includegraphics[width=8cm]{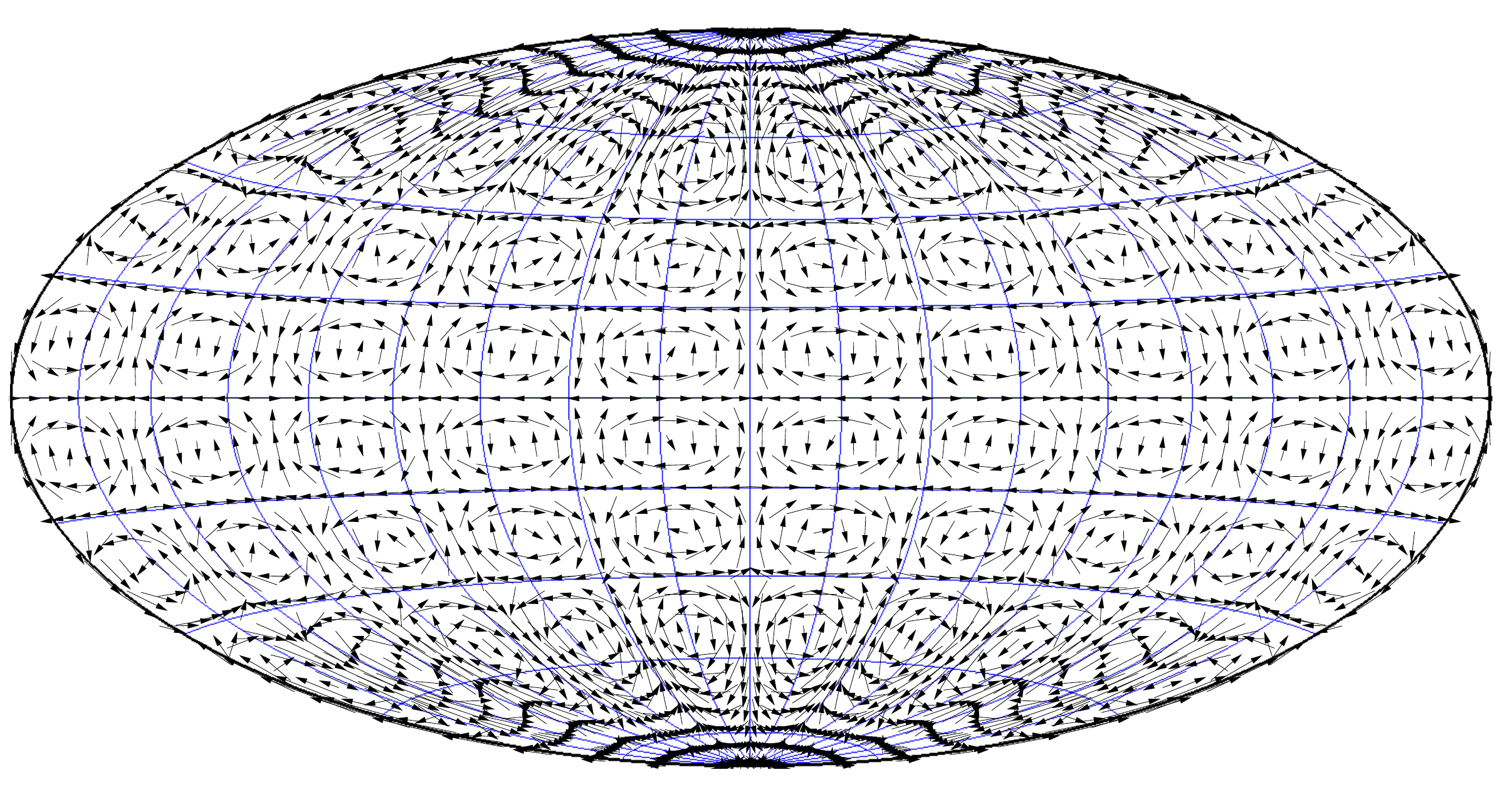}
}
\end{tabular}
\end{center}
 \caption{Examples of toroidal harmonics vector fields: $T_{1,1}$,
  $T_{2,1}$, $T_{5,3}$, $T_{10,5}$.  One sees the change in the
  angular resolution with higher degrees. The $T_{1,1}$ is a simple
  rotation about the $y$-axis.}
       \label{fig:toroidal_harm}
   \end{figure*}

\section{Transformation under rotation} \label{wigner}
\subsection{Overview}

A very important mathematical and practical feature of the expansion
of a vector field on the basis of the VSHs is how they transform under
a rotation of the reference axes. In short, when a vector field is
given in a frame ${\cal S}$ and decomposed over the VSHs in this
frame, one gets the set of components $t_{lm}, s_{lm}$ in this
frame. Rotating the reference frame ${\cal S}$ (e.g. transforming from
equatorial coordinates to galactic coordinates) to the frame
$\overline{\cal S}$, one gets the transformed vector field that can be
fitted again on a set of VSHs defined in the frame $\overline{\cal
  S}$.  This results in a new set of components $\overline{t}_{lm}$
and $\overline{s}_{lm}$, representing the vector field in this second
frame. Significant additional work may be required to carry out the
whole transformation of the initial field and perform the fit anew in
the rotated frame. Fortunately due to the narrow relationship between
VSHs and group representation, there is in fact a dramatic shortcut to
this heavy procedure that adds considerably to the interest of using
the VSH expansions in astronomy. This is illustrated in the
self-explanatory accompanying commutative diagram, showing that to a
space rotation (in the usual three-dimensional space)
$R$ corresponds to an operator
$\mathcal{R}$ acting in the space of the VSHs and allows
transforming $t_{lm}$ and $s_{lm}$ given in the initial frame into an
equivalent set in the rotated frame:
$$
\begin{CD}
{\cal S} @>R>> \overline{\cal S}\\
@V{VSH}VV @VV{VSH}V\\
\{t_{lm},s_{lm}\} @>\mathcal{R}>> \{\overline{t}_{lm},\overline{s}_{lm}\}
\end{CD}
$$
\noindent
The amount of computation is negligible since this is a linear
transformation between the coefficients of a given degree
$l$. Overall, the set of $t_{lm}$ and $s_{lm}$ is transformed into
itself without mixing coefficients of different degrees.

This operator $\mathcal{R}$ has been introduced by E. Wigner in 1927
as the D-matrices. A convenient reference regarding its definition and
properties is the book of \citet{VarshalovichMoskalevKhersonski1988}.
Section 7.3 of that book is specifically devoted to this topic.  The
Wigner matrices have been used for decades in quantum mechanics, but
probably they are not so well known in fundamental astronomy. We
confine ourselves here to a quick introduction of the main formulas
and conventions used in this paper. All the mathematical formulas have
been implemented in computer programs (independently in FORTRAN and
{\it Mathematica}), which can be requested from the authors.

For practical astronomical applications, there is an unexpected
benefit in this transformation under space rotation,
which is not shared by the
separate analysis on spherical harmonics of the components like
$\Delta \alpha\cos\delta$ and $\Delta \delta$, or their equivalent for
the proper motions. In general the two components for a given source
obtained from observations do not have the same accuracy and are
correlated. In the case of space astrometry missions like Hipparcos or
Gaia, the correlations are generally smaller in the ecliptic frame due
to the symmetry of the scanning in this frame compared to the
correlations in the equatorial frame, where the solution is naturally
available. Performing a least squares adjustment to compute $t_{lm}$
and $s_{lm}$ with correlated observations is a complication and
standard pieces of software often allow only for diagonal weight
matrix. One can easily get round this difficulty by carrying out the
fitting in a frame where the correlations can be neglected (assuming
such a frame does exist), and then rotate the coefficients with the
Wigner matrix into the equatorial frame. This is equivalent to
rigorously performing the fit in the equatorial coordinates with
correlated observations. On the other hand, by fitting the components with
scalar spherical harmonics, it is cumbersome to take the
correlations between the components properly into account, and the
coefficients in one frame do not transform simply into their
equivalent into a rotated frame.

More generally, if we allow for the correlations in the frame where the
least squares expansion over the VSHs is carried out, one can still
apply the Wigner rotation to the results and obtain the expansion in a
second frame. The results would be exactly the same as if a new least
squares  fit had been performed in this frame by rigorously
propagating the non-diagonal covariance matrix of the
observations to this second frame to generate the new weight
matrix. Using the Wigner matrix in this context is conceptually much
simpler and in keeping with the underlying group properties of the
VSHs, even though there might be no definite advantage for numerical
efficiency (but no disadvantage as well).

\subsection{Mathematical details}

Consider two rectangular Cartesian coordinate
systems ${\cal S}$ and $\overline{\cal S}$ of the same handedness.
Two coordinate systems are related by a rotation parametrised by three
Euler angles $a$, $b$, and $c$. For a given vector with components
$\ve{x}$ in ${\cal S}$, the components $\overline{\ve{x}}$ in
$\overline{\cal S}$ read as
\begin{equation}
\label{rotation-def}
\overline{\ve{x}} ={\bf R}_z(c)\,{\bf R}_y(b)\,{\bf R}_z(a)\,\ve{x},
\end{equation}
\noindent
where ${\bf R}_z$ and ${\bf R}_y$ are the usual matrix of (passive) rotation,
respectively, about axes $z$ and $y$
\begin{eqnarray}
\label{rotation-z}
{\bf R}_z(\varepsilon)&=&
\pmatrix{\cos\varepsilon& \sin\varepsilon&0\cr
-\sin\varepsilon&\cos\varepsilon&0\cr
0&0&1},
\end{eqnarray}
\begin{eqnarray}
\label{rotation-y}
{\bf R}_y(\varepsilon)&=&
\pmatrix{\cos\varepsilon&0&-\sin\varepsilon\cr
0&1&0\cr
\sin\varepsilon&0&\cos\varepsilon\cr}\,,
\end{eqnarray}
\noindent
where the order of arguments in the rotational matrices in
(\ref{rotation-def}) must be exactly followed. The inverse
transformation reads
\begin{equation}
\label{rotation-def-inverse}
\ve{x}={\bf R}_z(-a)\,{\bf R}_y(-b)\,{\bf R}_z(-c)\,\overline{\ve{x}}\,.
\end{equation}

Here the convention $z-y-z$ is used for the sequence of rotations
instead of the usual $z-x-z$ that is more common in dynamics and celestial
mechanics. The former choice has the great advantage of giving a
real matrix $d^l_{mn}(b)$ (see below) for the intermediate rotation,
and this convention is universally used in quantum mechanics and group
representation. The values of $a$, $b$, and $c$ for transformations
between the three usual frames in astronomy are listed in
Table.~\ref{tab:angles}. The relation between the $z-y-z$ and $z-x-z$
conventions is given by
\begin{equation}
\label{323-313}
{\bf R}_z(c)\,{\bf R}_y(b)\,{\bf R}_z(a)={\bf R}_z\left(c-\pi/2\right)\,{\bf R}_x(b)\,{\bf R}_z\left(a+\pi/2\right),
\end{equation}
\noindent
where ${\bf R}_x$ are the usual matrix of rotation
about axis $x$:
\begin{eqnarray}
\label{rotation-x}
{\bf R}_x(\varepsilon)&=&
\pmatrix{1&0&0\cr
0&\cos\varepsilon&\sin\varepsilon\cr
0&-\sin\varepsilon&\cos\varepsilon\cr}\,.
\end{eqnarray}

Then the transformations between the scalar and vector spherical
functions defined in ${\cal S}$ and $\overline{\cal S}$ are similar to
each other and read as
\begin{eqnarray}
\label{transformation-Y}
\overline{Y}_{lm}(\overline{\alpha},\overline{\delta})&=&
\sum_{m^\prime=- l}^l
D^l_{m^\prime m}(a,b,c)\,Y_{lm^\prime}(\alpha,\delta)\,,
\\
\label{transformation-T}
\overline{\ve{T}}_{lm}(\overline{\alpha},\overline{\delta})&=&
\sum_{m^\prime=- l}^l
D^l_{m^\prime m}(a,b,c)\,\ve{T}_{lm^\prime}(\alpha,\delta)\,,
\\
\label{transformation-S}
\overline{\ve{S}}_{lm}(\overline{\alpha},\overline{\delta})&=&
\sum_{m^\prime=- l}^l
D^l_{m^\prime m}(a,b,c)\,\ve{S}_{lm^\prime}(\alpha,\delta)\,,
\end{eqnarray}
\noindent
where $\overline{\alpha}$ and $\overline{\delta}$ are angles defined
as right ascension and declination (generally speaking, the longitude
and latitude of the spherical coordinates) from the components
$\overline{\ve{x}}$, while $\alpha$ and $\delta$ are those derived
from the components $\ve{x}$. In both cases Eq. (\ref{ni-alpha-delta})
is used.  Here $D^l_{mn}(a,b,c)$, $|m|\le l$, $|n|\le l$ are the
Wigner $D$-matrices (generalised spherical functions) defined as
\begin{eqnarray} \label{bigdlm}
D^l_{mn}(a,b,c)&=&e^{-\mi\left(m\,a+n\,c\right)}\,d^l_{mn}(b),
\end{eqnarray}
\begin{eqnarray}
\label{smalldlm}
d^l_{mn}(b)&=&(-1)^{m-n}\sqrt{(l+m)!(l-m)!(l+n)!(l-n)!}\,
\nonumber\\
&&\times
\sum_{k=k_{\rm min}}^{k_{\rm max}}(-1)^k
{{\left(\cos{b\over2}\right)}^{2l-2k-m+n}{\left(\sin{b\over2}\right)}^{2k+m-n}
\over k!(l-m-k)!(l+n-k)!(m-n+k)!}\,,
\nonumber\\[8pt]
&&
k_{\rm min}=\max(0,n-m),\quad k_{\rm max}=\min(l-m,l+n)\,.
\end{eqnarray}
\noindent
Many efficient ways to evaluate $D^l_{mn}(a,b,c)$ and $d^l_{mn}(b)$
can be found in relevant textbooks. For example, Section 4.21 of
\citet{VarshalovichMoskalevKhersonski1988} gives the expression of
$d^l_{mn}(b)$ in the form of a simple Fourier polynomial
\begin{equation}
d^l_{mn}(b)=\left(\mi\right)^{m-n} \sum_{k=-l}^l
d^l_{km}\left({\pi\over2}\right)\,
d^l_{kn}\left({\pi\over2}\right)\,e^{-\mi\,k\,b}\,,
\end{equation}
\noindent
where $d^l_{mn}\left({\pi\over2}\right)$ can be precomputed (using
the obvious simplification of (\ref{smalldlm})) as constants and used in
further calculations for a given rotational angle $b$. However, even a
straight implementation of (\ref{bigdlm})--(\ref{smalldlm}) does the
job without any problem for low to medium degrees.

The inverse transformations of the spherical functions read as
\begin{eqnarray}
\label{transformation-Y-inverse}
Y_{lm}(\alpha,\delta)&=&
\sum_{m^\prime=- l}^l
D^{l\,\ast}_{m\,m^\prime}(a,b,c)\,
\overline{Y}_{lm^\prime}(\overline{\alpha},\overline{\delta})\,,
\\
\label{transformation-T-inverse}
\ve{T}_{lm}(\alpha,\delta)
&=&
\sum_{m^\prime=- l}^l
D^{l\,\ast}_{m\,m^\prime}(a,b,c)\,
\overline{\ve{T}}_{lm^\prime}(\overline{\alpha},\overline{\delta})
\,,
\\
\label{transformation-S-inverse}
\ve{S}_{lm}(\alpha,\delta)
&=&
\sum_{m^\prime=- l}^l
D^{l\,\ast}_{m\,m^\prime}(a,b,c)\,
\overline{\ve{S}}_{lm^\prime}(\overline{\alpha},\overline{\delta})
\,,
\end{eqnarray}
\noindent
where the superscript `$\ast$' denotes complex conjugation.

Now what matters for our purpose is the transformation of the VSH
coefficients $t_{lm}$ and $s_{lm}$ as in (\ref{Vexpand}) under a rotation of
the coordinate system, and, more generally, that of coefficients
$f_{lm}$ in the case of an expansion of a scalar field in the scalar
spherical harmonics as given by (\ref{f-Ylm}).

Given the completeness of the scalar and vector basis functions, we
have defined the following expansions for scalar field $f$ and vector
field $\ve{V}$ on sphere (cf. (\ref{f-Ylm}) and (\ref{Vexpand})):
\begin{eqnarray}
\label{two-Y-expansions}
f&=&\sum_{l=0}^\infty\sum_{m=-l}^lf_{lm}\,Y_{lm}
=\sum_{l=0}^\infty\sum_{m=-l}^l\overline{f}_{lm}\,\overline{Y}_{lm}\,,
\\
\label{two-VSH-expansions}
\ve{V}&=&\sum_{l=1}^\infty\sum_{m=-l}^l
\left(t_{lm}\,\ve{T}_{lm}+s_{lm}\,\ve{S}_{lm}\right)
=\sum_{l=1}^\infty\sum_{m=-l}^l
\left(\overline{t}_{lm}\,\overline{\ve{T}}_{lm}+\overline{s}_{lm}\,\overline{\ve{S}}_{lm}\right).
\end{eqnarray}
\noindent
Substituting (\ref{transformation-Y})--(\ref{transformation-S}) or
(\ref{transformation-Y-inverse})--(\ref{transformation-S-inverse})
into (\ref{two-Y-expansions})--(\ref{two-VSH-expansions}) and changing
the order of summations, one gets the important transformation rules
for the coefficients
\begin{eqnarray}
\label{transformation-flm}
f_{lm}&=&\sum_{m^\prime=- l}^l D^l_{m\,m^\prime}(a,b,c)\,\overline{f}_{lm^\prime}\,,\label{direct1}
\\
\label{transformation-tlm}
t_{lm}&=&\sum_{m^\prime=- l}^l D^l_{m\,m^\prime}(a,b,c)\,\overline{t}_{lm^\prime}\,,\label{direct2}
\\
\label{transformation-slm}
s_{lm}&=&\sum_{m^\prime=- l}^l D^l_{m\,m^\prime}(a,b,c)\,\overline{s}_{lm^\prime}\,,\label{direct3}
\end{eqnarray}
\noindent
and the inverse transformations
\begin{eqnarray}
\label{transformation-flm-inverse}
\overline{f}_{lm}&=&\sum_{m^\prime=- l}^l D^{l\,\ast}_{m^\prime m}(a,b,c)\,f_{lm^\prime}\,,\label{inverse1}
\\
\label{transformation-tlm-inverse}
\overline{t}_{lm}&=&\sum_{m^\prime=- l}^l D^{l\,\ast}_{m^\prime m}(a,b,c)\,t_{lm^\prime}\,,\label{inverse2}
\\
\label{transformation-slm-inverse}
\overline{s}_{lm}&=&\sum_{m^\prime=- l}^l D^{l\,\ast}_{m^\prime m}(a,b,c)\,s_{lm^\prime}\,.\label{inverse3}
\end{eqnarray}

\begin{table}[ht!]
\setlength{\tabcolsep}{3mm}
{\small
\begin{center}
\caption{Euler angles $a$, $b$ and $c$ applicable to the
change of astronomical frame.
The obliquity has been taken for J2000 and equatorial frame can be seen as
the same as ICRF at this level of accuracy. }
\vspace{2mm}
  \begin{tabular}{llll}
   \hline \\[-6pt]
   Transformation  &$a$  & $b$ & $c$ \\[0mm]
                   & deg  &  deg &   deg \\
   \hline\\[-4pt]
Equatorial to Ecliptic  &    270.0     &  23.4393 &  90.0\\
Equatorial to Galactic  &    192.8595  &  62.8718 &  57.0681 \\
Ecliptic to Galactic    &    180.0232  &  60.1886 &  83.6160\\
\hline
  \end{tabular}
  \label{tab:angles}
\end{center}
}
\end{table}

Once the expansion has been computed in one frame, it is very easy to
compute the expansion in any other frame related by a rotation from
the initial frame.  Numerical checks have been performed with the
analysis of a vector field in two frames, followed by the
transformations (direct and inverse) of the coefficients $t_{lm}$ and
$s_{lm})$ with (\ref{direct1})--(\ref{direct3}) and
(\ref{inverse1})--(\ref{inverse3}).

These transformation rules allow us also to prove the invariance of
the Euclidean norm of the set of coefficients of a given degree:
\begin{eqnarray}
\label{P_l^Y}
{\cal P}^Y_{l}&=&\sum_{m=- l}^l f_{lm}\,f^\ast_{lm}
=\sum_{m=- l}^l \overline{f}_{lm}\,\overline{f}^{\,\ast}_{lm},
\\
\label{P_l^t}
{\cal P}^t_{l}&=&\sum_{m=- l}^l t_{lm}\,t^\ast_{lm}
=\sum_{m=- l}^l \overline{t}_{lm}\,\overline{t}^{\,\ast}_{lm},
\\
\label{P_l^s}
{\cal P}^s_{l}&=&\sum_{m=- l}^l s_{lm}\,s^\ast_{lm}
=\sum_{m=- l}^l \overline{s}_{lm}\,\overline{s}^{\,\ast}_{lm}.
\end{eqnarray}
\noindent
This can be easily demonstrated by substituting
(\ref{transformation-flm})--(\ref{transformation-slm}) into corresponding
equation and using the well-known relation
\begin{equation}\label{unitary}
\sum_{m=-l}^l D^l_{m\,m^\prime}(a,b,c)\,D^{l\,\ast}_{m\,m^{\prime\prime}}(a,b,c)
=\delta_{m^\prime m^{\prime\prime}}\,,
\end{equation}
\noindent
which is a special case of the general addition theorem for the Wigner
$D$-matrices \citep[][Section
4.7]{VarshalovichMoskalevKhersonski1988}.  The invariance of ${\cal P}_l$
will be used in Section~\ref{sect-power} to assess the level of
significance of the coefficients.

\section{Relation of the first-order VSH to global features}
\label{sect:global}

The multipole (VSH) representation of a vector field has some
similarity with a Fourier or a wavelet decomposition since as the
degree $l$ increases, one sees smaller details in the structure of the
field. In the case of the comparison between two catalogues,
increasing the degree reveals the zonal errors of the catalogue, while
the very first degrees, say $l=1$ and $l=2$, show features with the
longer ``wavelengths''. In particular, the first degree both in the
toroidal and spheroidal harmonics is linked to very global features,
such as rotation between the two catalogues or a systematic dipole
displacement, like a flow from a source to a sink located at the two
poles of an axis. It must be stressed that these global features are
found in a blind way, without searching for them.  They come out
naturally in the decomposition as the features having the least
angular resolution, and they are interpreted as a rotation or a dipole
glide a posteriori.

\subsection{Rotation}

Consider a infinitesimal rotation on a sphere given by an the rotation
vector $\vec{R}$ whose components in a particular frame are $(R_1,
R_2, R_3)$. This rotation generates the vector field
\begin{equation}\label{omegarot}
    \vec{V}^R = \vec{R}\times\vec{u}\,,
\end{equation}
\noindent
where $\vec{u}$ is the unit radial vector given by (\ref{ni-alpha-delta}).
Equation~(\ref{omegarot}) can be projected on
the local tangent plane and leads to
 \begin{eqnarray}\label{rotfield}
   V_\alpha^R &=& \ve{V}^R\cdot\ve{e}_\alpha=
-R_1 \sin\delta\, \cos\alpha -R_2\sin\delta\,\sin\alpha
+R_3\cos\delta\,, \\
   V_\delta^R &=&\ve{V}^R\cdot\ve{e}_\delta=
R_1 \sin\alpha -R_2\cos\alpha\,.
\end{eqnarray}
\noindent
From the explicit expressions of the toroidal harmonics given in
Table~\ref{tab:Tharm}, one sees that any infinitesimal rotation field
has non-zero projection only on the $\vec{T}_{1m}$ harmonics, and is
orthogonal to any other set of toroidal or spheroidal harmonics.
Therefore (\ref{omegarot}) can be written as a linear combination of the $\vec{T}_{1m}$.
One has explicitly the equivalence
\begin{equation}\label{rottoroidal}
    \vec{V}^R = t^R_{10}\vec{T}_{10} + t^R_{11}\vec{T}_{11} +t^R_{1,-1}\vec{T}_{1,-1}
\end{equation}
\noindent
where
\begin{eqnarray}
\label{t10-rotation}
  t_{10}^R &=&  \sqrt{\frac{8\pi}{3}}\ R_3\,, \\
\label{t11-rotation}
  t_{11}^R = -t^{R\,\ast}_{1,-1}&=& \sqrt{\frac{4\pi}{3}}\ (-R_1 + \mi\,R_2)\,,
\end{eqnarray}
\noindent
allowing the three parameters $\ve{R}$ of the rotation to be extracted
from the harmonic expansion of degree $l=1$. A change of reference
frame will lead to the same rotation vector expressed in the new
reference frame. Given the importance of a rotation between two
catalogues or in the error distribution of a full sky astrometric
catalogue, having the rotational vector field as one of the base
functions is a very appealing feature of the VSH. This contrasts with
the analysis of each components of the vector on scalar spherical
harmonics, in which a plain rotation projects on harmonics of any
degree as shown in \citet{Vityazev1997,Vityazev2010} or in Appendix
\ref{annex_2}. However, the analysis of each component as scalar
fields retains its interest if one has good reasons to assume that the
field has been generated by a process in which the two spherical
coordinates have been handled more or less separately. In this case
there is even no reason to change the frame for another one since the
two components may have very different statistical properties.

\subsection{Glide}

The rotation field is the most common fully global signature on a
spherical vector field, and often it even seems to be the only one
deserving the qualification of global. However, if we consider a
rotational field $\vec{G} \times\ve{u}$, one may draw at each point on
the sphere a vector perpendicular to the rotation field and lying in
the tangent plane to the sphere. In this way we generate another field
$\ve{V}^G$ with axial symmetry, fully orthogonal to the rotation field
$\vec{G} \times\ve{u}$. This new field has no projection on the
$\vec{T}_{1m}$ harmonics and, in fact, on no other $\vec{T}_{lm}$. Its
components are given by
\begin{eqnarray}\label{glidefield}
V_\alpha^G &=& -G_1 \sin\alpha +G_2\cos\alpha\,, \\
V_\delta^G &=& -G_1 \sin\delta\,\cos\alpha -G_2\sin\delta\,\sin\alpha+G_3\cos\delta\,,
\end{eqnarray}
\noindent
where the vector $\vec{G} =(G_1, G_2, G_3)$ determines the orientation
and the magnitude of the field, as $\vec{R}$ did for the rotational
field. Going to Table~\ref{tab:Sharm} it is easy to see that it is
equivalent to the $\vec{S}_{1m}$ field with the decomposition
\begin{equation}\label{rotspheroidal}
    \vec{V}^G = s^G_{10}\vec{S}_{10} + s^G_{11}\vec{S}_{11} +s^G_{1,-1}\vec{S}_{1,-1}
\end{equation}
\noindent
where
\begin{eqnarray}
  s_{10}^G &=&  \sqrt{\frac{8\pi}{3}}\,G_3\,, \\
  s_{11}^G &=&  \sqrt{\frac{4\pi}{3}}\,(-G_1 + \mi\,G_2)\,,
\end{eqnarray}
\noindent
showing its close relationship with a rotation field. A corresponding
picture can be seen in Fig.~\ref{fig:spheroidal_harm} for an axis along the $y$-axis.

The same can be stated somewhat differently: we have introduced the
glide\footnote{We coined this word to have something as easy
to use as \textsl{rotation} to name the transformation and the
feature and to convey in a word the idea of a smooth flow between
the two poles.} vector field $\vec{V}^G$, which is orthogonal to both
radial vector $\ve{u}$ and the rotation field $\vec{G}\times\vec{u}$,
or equivalently one has
\begin{equation}\label{glide}
   \vec{V}^G = \vec{u}\times(\vec{G}\times\vec{u}) =  \vec{G} - (\vec{G}\cdot\vec{u})\,\vec{u}\,,
\end{equation}
\noindent
which shows clearly that the associated vector field is just the
projection of $\vec{G} $ on the surface of the unit sphere (we have
above $\vec{G}$ minus its radial component), that is to say the projection
of an axial, or dipole, field on the sphere.

As a pattern, a glide field is as global a feature as a rotation, and
can be seen as a regular flow between a source and a sink
diametrically opposed. From the astronomical point of view this is a
field associated to a motion of the observer toward an apex, with all
the stars showing a kinematical stream in the opposite direction. For
extragalactic sources like quasars, this also characterises the effect
induced on QSO systematic proper motions resulting from the
acceleration of the observers with respect to the frame where the QSOs
are at rest on average \citep{Fanselow1983,SoversFanselowJacobs1998,
kova2003,kopeikin2006,TitovEtAl2011}. The main source of this
acceleration  is thought to be the centripetal acceleration from the
galactic rotation. Its unambiguous determination should be
achieved with Gaia. It is important to recognise that in the general
analysis of catalogues or spherical vector fields, rotation is not
more fundamental a feature than glide, and both must be searched for
simultaneously. Otherwise, for most of the distributions of the data
on the sphere, when the discrete orthogonality conditions between the
VSHs are not fully satisfied, there will be a projection of the
element not included in the model in the subspace generated by the
other one. Thus, fitting only the rotation, the rotation vector will be
biased if there is a glide not accounted for in the data model. And
reciprocally for the glide without rotation properly fitted.

  \begin{figure*}[htb]
\begin{center}
\begin{tabular}{cc}
\subfigure[Rotation]{
 \includegraphics[width=8cm]{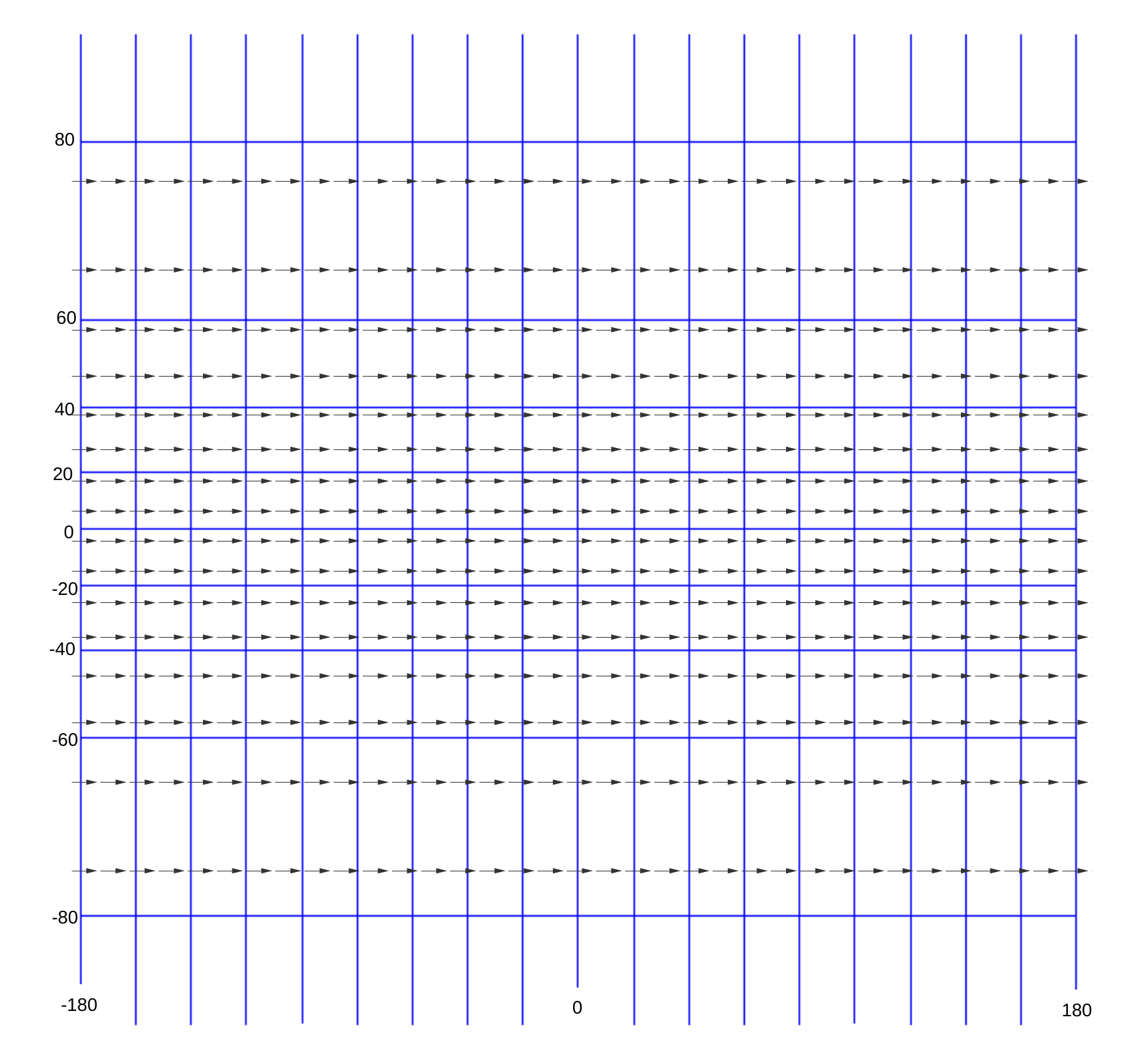}
}
&
\subfigure[Glide]{
 \includegraphics[width=8cm]{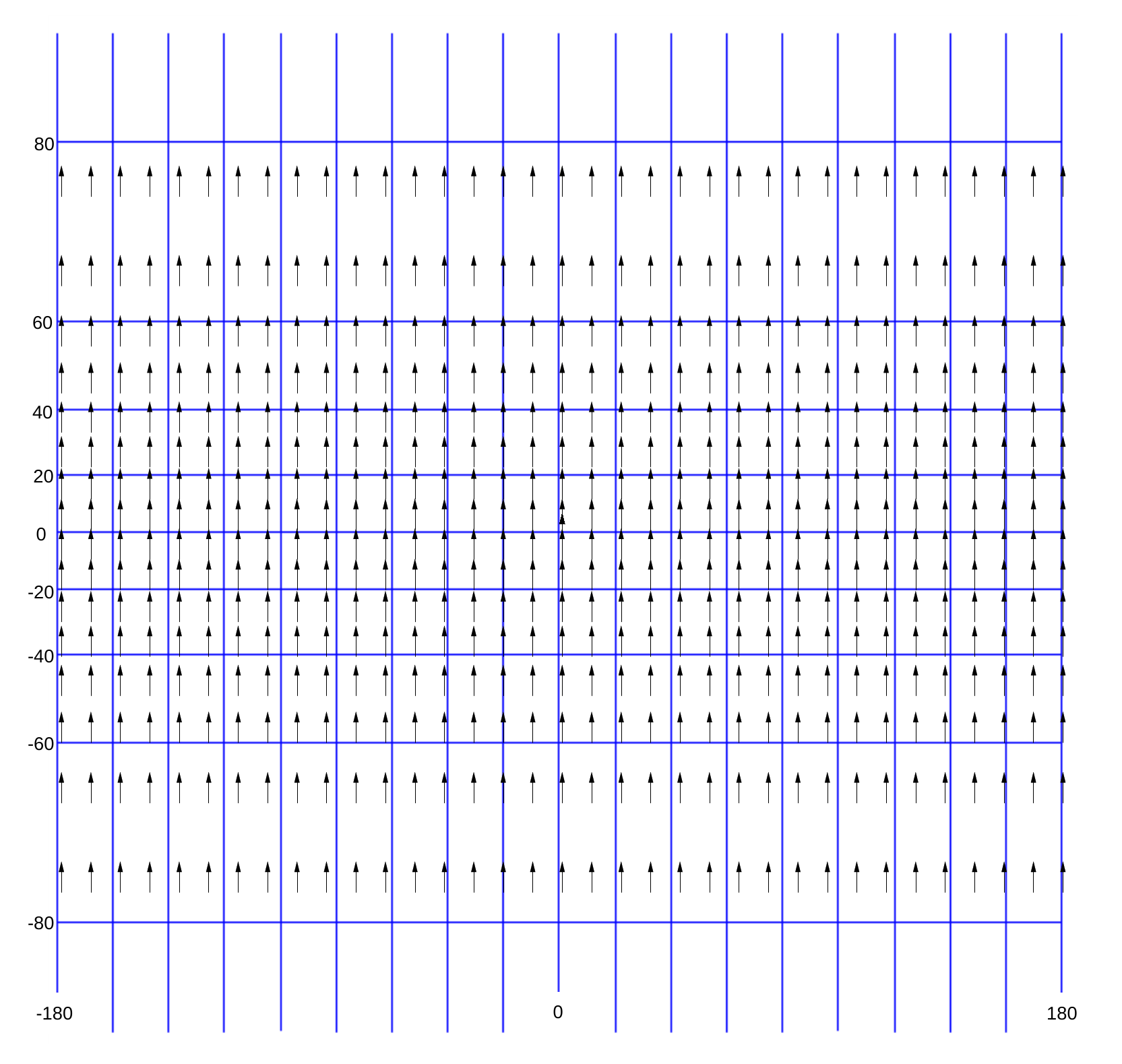}
}
\end{tabular}
\end{center}
 \caption{Rotation and glide with the $z$-axis
   shown here with a Mercator projection to emphasise the close
   relationship between the two vector fields. The constant
   displacement in longitude from the rotation has its counterpart
   with a constant glide displacement in the reduced latitude
   (Mercator latitude).}
\label{fig:rot_glide}
   \end{figure*}

\subsection{Duality of the rotation and glide effects}\label{duality}

Finally the duality between the infinitesimal rotation and glide
can be made striking if the fields are plotted in the Mercator
projection, as illustrated in Fig.~\ref{fig:rot_glide}. Both
effects are along the $z$-axis and have the same magnitude for
$\vec{R}$ and $\vec{G}$.  Thus they are rotation and glide
with the same axis. The effect of a small rotation is seen as
a uniform translation in longitude, identical to all latitudes, as
expected for a cylindrical projection. For the glide, the
translation is uniform in the reduced latitude given by the
Mercator transformation
\begin{equation}\label{mercator}
   \delta_\mathrm{M} = \ln\left(\tan\left(\frac{\delta}{2} + \frac{\pi}{4}\right)\right)
\end{equation}
\noindent
and does not depend on the longitude.

There is a clear similarity between the two fields (both represent a
mapping of the sphere onto itself with two fixed
points: the poles), but also an important difference seen with the
Mercator projection: for the rotations, the mapping of a flow line
(the continuous line tangent to the vectors of the field) is bound
between the two extreme boundaries $-\pi$ and $+\pi$, which must be
identified as the same line on the sphere. The rotation axis is the
natural source of a cylindrical symmetry shown by the Mercator
projection (or any other cylindrical projection). For the glide, it
goes very differently since the two end points of a flow line are sent
to infinity, in both directions, and can only be identified in the
projective space, a much more complex manifold than the Euclidean
cylinder.  Could this important topological difference be the
underlying reason why only the rotation is usually considered as the
only obvious global structure?  We do not have the answer and leave this
issue open for the moment. As long as infinitesimal transformations
are concerned, which is exactly what matters for the analysis of
astrometric catalogues, the two effects are in fact very similar and
must be taken into account together as a whole.

\section{Practical implementation}\label{implementation}

In principle, once the difference between two catalogues has been
formed or when a vector field on a sphere (e.g. a set of proper
motions) is available, it is normally trivial to project the
underlined vector field on the VSH, by computing the integrals
(\ref{coef}) with an appropriate numerical methods. This will work as
long as the distribution of the sources on the unit sphere preserves
the orthogonality conditions between the different base
functions. What is precisely meant is the following: using the same
numerical method to evaluate the Fourier integrals, one should check that
Eqs.~(\ref{TdotT}), (\ref{SdotS}) and (\ref{SdotT}) hold with a level
of accuracy compatible with the noise up to some degree $l_{\rm
  max}$. Unless one has many sources almost evenly distributed, this
condition is rarely fulfilled. However, the fact that the coefficients
can be expressed by integrals, that is to say by direct projection on
the base functions, is a simple consequence of the application of a
least squares criterion leading to a diagonal normal matrix when the
orthogonality condition holds. Starting one step back to the
underlying least squares principle we can directly fit the
coefficients to the observed components of the field with the model
(\ref{Vexpandreal}), by applying a weighting scheme based on the noise in the
data. In case the orthogonality conditions hold, the solution will be
equivalent to the evaluation of the integrals. When this assumption is
not true, meaning with the discrete sampling and/or uneven weight distribution
at the source coordinates, the harmonics are not numerically orthogonal, but
the least squares fitting provides an unbiased estimate of the coefficients
$t_{lm}$ and $s_{lm}$, together with their covariance matrix.

\subsection{Least squares solution}

The fitting model is given by (\ref{Vexpandreal}) up to the degree
$l=l_{\rm max}$. It is fitted to a set of $N$ data points sampled at ($\alpha_k$,
$\delta_k$), $k = 1,2,\cdots, N$, where at each point the two
components of the vector field $\vec{\Delta}$ (e.g., difference between two
catalogues, or proper motion components) are given as
\begin{equation}\label{vectorfield}
\vec{\Delta}(\alpha_k, \delta_k) =
\Delta^\alpha_k\,\ve{e}_\alpha + \Delta^\delta_k\,\ve{e}_\delta, \ k = 1, 2, \cdots, N\,,
\end{equation}
\noindent
together with the estimates of their associated random noise
$\sigma_{\Delta^\alpha_k}$ and $\sigma_{\Delta^\delta_k}$. For the
sake of simplicity we assume here that the covariance matrix
between the observations is diagonal, meaning the observations are  realisations of
independent random variables. The observations are fitted to the model
(\ref{Vexpandreal}) with the least squares criterion
\begin{equation}\label{lsqr}
   \min_{t_{lm}, s_{lm}} \sum_{k=1}^{N}  \left(\frac{\left(\Delta^\alpha_k-V_\alpha(\alpha_k,\delta_k)\right)^2}{\sigma_{\Delta^\alpha_k}^2} + \frac{\left(\Delta^\delta_k-V_\delta(\alpha_k,\delta_k)\right)^2}{\sigma_{\Delta^\delta_k}^2}\right)\,.
\end{equation}
\noindent
From (\ref{Vexpandreal}) and (\ref{Vexpandcomp}) we see that with $N$
observations one has $2N$ conditional equations to determine the
$2l_{\rm max}(l_{\rm max}+2)$ coefficients $(t_{l0},\,t^{\Re}_{lm},\,
t^{\Im}_{lm},\,s_{l0},\, s^{\Re}_{lm},\,s^{\Im}_{lm})$,\ $m =1,2,
\ldots, l$,\ $l = 1,2,\ldots, l_{\rm max}$. The size of the design matrix is
$2N\,\times\,2l_{\rm max}(l_{\rm max}+2)$, which can be very large in analysis
of astrometric surveys, with $N$ above $10^5-10^6$ and a sensitivity on a small enough
angular scale to represent the regional errors leading to explore up
to $ l_{\rm max}\sim 15-20$. The data storage could become
a problem if one wishes to store and access the full design
matrix. One solution to get round this problem is to build up the
normal matrix on the fly, taking its symmetry into account. For a large data set
($N\gg l_{\rm max}^2$),
this bookkeeping part will be the most demanding in terms of computation time.
For fixed $N$ the computational time for this part of calculations
grows like $l_{\rm max}^4$. In general the problem is
rather well conditioned, with small off-diagonal terms, and is not
prone to generate numerical problems to get the least squares
solution.

\subsection{Significance level}
\label{sect-power}

The least squares solution ends up with an estimate for each of the
coefficients included in the model. The model must go to a certain
maximum value of $l_{\rm max}$ by including all the component
harmonics of order $m$, for each $ l \le l_{\rm max}$. The total
number of unknowns is then $2l_{\rm max}(l_{\rm max}+2)$ and grows
quadratically with $l_{\rm max}$. Within a degree $l$, one can decide
on the significance of each coefficient of order $m$, from its
amplitude compared to the standard deviation.  However, these
individual amplitudes are not invariant by a change of reference
frame, like the components of a vector in ordinary space, and
the significance of a particular component is not preserved through
a rotation of the frame. On the other hand, the whole set of $t_{lm}$
or $s_{lm}$ for a given $l$ behaves like a vector under rotation,
and the significance (or lack of) of the vector itself, that is to
say its modulus, is intrinsic and remains true or false in any other
rotated frame. Therefore it is more interesting and more useful in
practice to investigate the significance level for each
degree $l$ rather than that of the individual coefficients.  The
reason behind this statement is deeply rooted in the invariance
property of the $Y_{lm}$ under a rotation in the three-dimensional
Euclidean space and the fact that the set of spherical harmonics
$Y_{lm}$ of degree $l$ is the basis of an irreducible representation
of the rotation group of the usual three-dimensional space. Therefore,
this set is globally invariant under a rotation, meaning there is
linear transformation linking the $Y_{lm}$ in one frame to their
equivalent of same degree $l$ in the other frame (see Section
\ref{wigner}).

Therefore, one should consider the $2l+1$ elements of the set
$Y_{lm}$, for $ m = -l, \dots l$ like the components of a vector
that are transformed into each other under a rotation, keeping the
Euclidean norm unchanged in the process. Given the linearity
of the derivative operator, this property extends naturally to the
$\vec{T}_{lm}$ and $\vec{S}_{lm}$. For the same reason, the
coefficients $t_{lm}$ and $s_{lm}$ in the expansion of a vector field
must be considered together within a degree $l$ and not examined
separately, unless one has good reasons to interpret physically one or
several coefficients in a particular frame: the Oort coefficients in
the Galactic frame are one of the possible examples, as shown in
\citet{Vityazev2010}.  More generally a wide range of parameters
in Galactic kinematics have very natural representations on vector
harmonics, such as the solar motion towards the apex or a more general
description of the differential rotation than the simple Oort's
modelling (see for example \citet{Mignard2000a}).

Therefore it is natural to look at the power of the vector field and
at how this power is spread over the different degrees $l$. Let
$\mathcal{P}$ be the power of a real vector field $\ve{V}(\alpha,
\delta)$ on a sphere defined by the surface integral,\footnote{The
  power is often defined with a normalisation factor $1/4\pi$ before
  the integral. The present choice leads to simpler expressions for
  the power in terms of the coefficients $t_{lm}$ and $s_{lm}$ and has
  been preferred.}
\begin{equation}\label{power}
  \mathcal{P}(\vec{V}) = \int_\Omega\, |\vec{V}|^2\, d\Omega\,,
\end{equation}
\noindent
or its discrete form used hereafter with numerical data,
\begin{equation}\label{powernum}
  \mathcal{P}(\vec{V}) \simeq   \frac{4\pi}{n}\sum_{i=1}^{n} |\vec{V}_i|^2\,,
\end{equation}
\noindent
where $n$ is the number of samples on the sphere.

When projected on the VSH one finds easily with (\ref{TdotT}),
(\ref{SdotS}), and (\ref{Vexpand}) that
\begin{eqnarray}\label{Parseval1}
   \mathcal{P}(\vec{V}) &=& \sum_{l=1}^{l_{\rm max}}\left(\mathcal{P}^t_l+\mathcal{P}^s_l\right)\,,
\nonumber
\\
\label{P^t_l}
\mathcal{P}^t_l&=&\sum_{m=-l}^{l} t_{lm}\,t^{\ast}_{lm}\,,
\nonumber
\\
\label{P^s_l}
\mathcal{P}^s_l&=&\sum_{m=-l}^{l} s_{lm}\,s^{\ast}_{lm}\,.
\end{eqnarray}
\noindent
For a real vector field having the symmetries (\ref{reim1})--(\ref{reim1-s})
one gets
\begin{eqnarray}\label{Parseval2}
\label{Parseval3}
\mathcal{P}^t_l&=&t_{l0}^2 +2 \sum_{m=1}^{l} |t_{lm}|^2=t_{l0}^2 +2 \sum_{m=1}^{l} \left((t_{lm}^{\Re})^2+(t_{lm}^{\Im})^2\right)\,,
\nonumber
\\
\mathcal{P}^s_l &=&s_{l0}^2 +2 \sum_{m=1}^{l} |s_{lm}|^2=  s_{l0}^2
+2 \sum_{m=1}^{l}\ \left(\,(s_{lm}^{\Re})^2+(s_{lm}^{\Im})^2 \right)\,.
\end{eqnarray}
\noindent
The quantities $\mathcal{P}^t_l$ and $\mathcal{P}^s_l$
are the powers of toroidal and spheroidal components of the degree $l$
of the field. Both quantities are
scalars invariant under rotation of the coordinate system
(see Section~\ref{wigner}).
Thus $\mathcal{P}^t_l$ and $\mathcal{P}^s_l$ represent an
intrinsic property of the field, similar to the
magnitude of a vector in an ordinary Euclidean vector space.

We can now take up the important issue of the significance level of
the subset of the coefficients $s_{lm}$ and $t_{lm}$ for each degree
$l$. In harmonic or spectral analysis, this is always a difficult
subject, because it needs to scale the power against some typical
value that would be produced by a pure noisy signal. In general one
can construct a well-defined test, with secure theoretical
foundations, but with assumptions about the time or space sampling,
which are rarely found in the real world. Therefore the theory
provides a good guideline, but not necessarily a fully safe solution
applicable in all situations. The key issue here is to construct a
criterion that is well understood from the statistical view point and
robust enough to be usable for mild departure from the underlying
assumptions, like regular time sampling in time series. We offer below
three possibilities for testing whether the power $\mathcal{P}^t_l$
and $\mathcal{P}^s_l$ is significant and therefore that there is a
true signature of the VSH of degree $l$ in the vector field. In the
following we drop superscripts `$s$' and `$t$' and write simply
$\mathcal{P}_l$.  Relevant formulas below are valid separately for
$\mathcal{P}^t_l$ and $\mathcal{P}^s_l$ or for
$\mathcal{P}^t_l+\mathcal{P}^s_l$.  The involved coefficients will be
denoted as $r_{lm}$ to represent either $t_{lm}$ or $s_{lm}$ or both.

Start from the null hypothesis $H_0$ that the signal is just made of
noise, producing some values of $\mathcal{P}_l$ in the analysis.  One
needs to build a test to tell us that above a certain threshold, the
value of $\mathcal{P}_l$ is significant and would only be produced by
pure noise with a probability $p \ll 1$. To build the test properly,
the probability density function (PDF) of $\mathcal{P}_l$ is
required. It can be obtained with the additional assumptions that the
sources are rather evenly distributed on the sphere. The least squares
fitting also gives us the variance-covariance matrix of the unknowns
$s_{lm}$ and $t_{lm}$. Given the orthogonality properties of the VSH,
the normal matrix is almost diagonal if the distribution of sources
over the sky and the associated errors of the quantity representing
the vector field are homogeneous. In this case the covariance matrix is
also almost diagonal.

Consider the coefficients $r_{l0}$,
$r_{lm}^{\Re}$, and $r_{lm}^{\Im}$ in (\ref{Vexpandreal}) with a fixed
degree $l$.  Each diagonal term of the relevant part of the
normal matrix (unweighted here) takes the form
\begin{eqnarray}\label{normmatrix}
 \phantom{4}\sum R_{l0}^2  &\mathrm{\hspace{1cm} for \hspace{1cm} }& r_{l0}\,, \\
 4\sum (R_{lm}^{\Re})^2    &\mathrm{\hspace{1cm} for \hspace{1cm} }&  r_{lm}^{\Re}\,,\\
  4\sum (R_{lm}^{\Im})^2   &\mathrm{\hspace{1cm} for \hspace{1cm} }&  r_{lm}^{\Im}\,,
\end{eqnarray}
\noindent
where the sum extends over all the sources on the sphere. Here
$R_{lm}$ stands for either $T_{lm}$ or $S_{lm}$. The first term with
$m=0$ does not depend on the right ascension $\alpha$, while the others
have a part in $\cos^2\!m\alpha$ and $\sin^2\!m\alpha$, whose average
is $1/2$ for $\alpha$ ranging from $0$ to $2\pi$ and a relatively
regular distribution in $\alpha$. Given the normalisation of the VSHs,
we now see that the diagonal terms with $m\neq 0$ are just two times the
term with $m=0$. The covariance matrix being the inverse of the normal
matrix, one eventually has the variances of the unknowns as
\begin{eqnarray}\label{variances}
\mathrm{var}( r_{l0}) &=& \sigma^2_{l0}\,, \label{variances1} \\
\mathrm{var}( r_{lm}^{\Re}) &=& \sigma^2_{l0}/2\,, \label{variances2}\\
\mathrm{var}( r_{lm}^{\Im}) &=& \sigma^2_{l0}/2\,, \label{variances3}
\end{eqnarray}
where $\sigma^2_{l0}$ is the variance of $r_{l0}$. Then dividing
$\mathcal{P}_l $ in (\ref{Parseval3}) by
$\sigma^2_{l0}$ leads to the normalised power:
\begin{equation}\label{Parsevalchi}
 W_l = \frac{\mathcal{P}_l}{\sigma^2_{l0}} =
\frac {r_{l0}^2}{\sigma^2_{l0}}
+ \sum_{m=1}^{l}\,\left( \frac{(r_{lm}^{\Re})^2}{\sigma^2_{l0}/2}
+\frac{(r_{lm}^{\Im})^2}{\sigma^2_{l0}/2}\right).
\end{equation}
With (\ref{variances2})--(\ref{variances3}) and the assumption that
the signal is pure white Gaussian noise, this is the sum of $2l+1$
squares of standard normal random variables with zero mean and unit
variance. It is well known that $W_l$ follows a $\chi^2$ distribution
with $n= 2l+1$ degrees of freedom.  With this normalisation, $W_l$ is
strictly proportional to the power $\mathcal{P}_l$.

Under the $H_0$ hypothesis that the signal is just white noise, one
can build a  one-sided test to decide whether a harmonic of degree $l$ is
significant with the probability level $\gamma$:
\begin{equation}\label{signifchi2}
    P(W_l > w) = 1 - \int_0^w \chi^2_{2l+1}(x) dx ={\Gamma(l+1/2,w/2)\over\Gamma(l+1/2)}< \gamma \,,
\end{equation}
\noindent
where $\Gamma(a,x)$ and $\Gamma(a)$ are the incomplete and complete
gamma functions, respectively. This quantity can be easily computed.
In some cases it may be easier to use the transformation of
\citet{wilson31},
\begin{equation}\label{wilson}
  Z = \sqrt{\frac{9n}{2}}
\left[\,\left(\frac{W_l}{n}\right)^{1/3} - \left(1- \frac{2}{9n}\right)\,\right]\,,
\end{equation}
where $n$ is the number of degrees of freedom.  It is well known that $Z$
approximately follows,
even for small $n$, a standard normal distribution of zero mean and
unit variance. A test risk of $\gamma =0.01$ is achieved if $Z >
2.33$ or at the level of $\gamma =0.025$ is $Z > 1.96$. Experiments on
simulated data have shown that the key elements in the above
derivation, namely that $\sigma_{l0} \simeq \sqrt{2}\,\sigma_{lm}, m >0$
is practically achieved when sources are rather regularly distributed
in longitudes, even  with irregularities in latitudes. The reason
is basically that this factor $\sqrt{2}$ follows from the average of
$\cos^2 m\alpha$ and $\sin^2 m\alpha$, while the latitude effects are
shared in a similar way in all the $r_{lm}$.

Alternatively, instead of scaling the power by only $\sigma^2_{l0}$ and
taking advantage of the asymptotic properties
(\ref{variances1})--(\ref{variances3}), it would have been more natural
to define a reduced power with
\begin{equation}\label{Parsevalchi2}
 \widetilde{W_l} =  {\left(\frac{r_{l0}}{\sigma_{l0}}\right)}^2
+ \sum_{m=1}^{l}\ \left({\left(\frac{r_{lm}^{\Re}}{\sigma_{lm}^{\Re}}\right)}^2
+{\left(\frac{r_{lm}^{\Im}}{\sigma_{lm}^{\Im}}\right)}^2\right)\,,
\end{equation}
\noindent
which is not strictly proportional to the power. But what matters in
practice is whether one can construct an indicator, related to the
power, with well defined statistical properties, so one may relax the
constraint of having exactly the power, provided that a reliable test of
significance can be built. For least squares solutions with Gaussian
noise, and small correlations between the estimates of the unknowns,
it is known that each coefficient follows a normal distribution of
zero mean whose variances can be estimated with the inverse of the
normal matrix. Therefore, $\widetilde{W_l}$ is also a sum of squares of
standard normal variables and its PDF is that of a $\chi^2$ distribution with
$2l+1$ degrees of freedom. For regular distributions of the data
points, the two reduced powers ${W_l}$ and $\widetilde{W_l}$ are equivalent. So
either test should give the same result. In
(\ref{Parsevalchi2}) the multiplying factor two before the sum in
(\ref{Parseval3}) has been absorbed by $\mathrm{var}(\sqrt{2}Y) =
2\,\mathrm{var}(Y)$ (i.e. by $\sigma_{lm}^{\Re}$ and $\sigma_{lm}^{\Im}$ being
approximately a factor $\sqrt{2}$ smaller than $\sigma_{l0}$).

Finally, we can also go back to the first test in
(\ref{Parsevalchi}), which is the power scaled by the variance of
the component $s_{l0}$. Since the variance ${\sigma^2_{l0}}$ is in
practice an estimate of a random variable based on the
data, it may happen that the value produced by the least squares is
unusually small, which will make ${W_l}$ too large and trigger a false
detection. Therefore, while this test has a good statistical foundation
for regular distribution of sources, it is not very robust to
departure from this assumption. We suggest improving this robustness, a
very desirable feature in the real data processing, by replacing the
scaling factor $\sigma^2_{l0}$ by the average of the set
$\left(\,\sigma^2_{l0}, 2(\sigma_{lm}^{\Re})^2, 2(\sigma_{lm}^{\Im})^2\,\right)$,\ $m
=1, \ldots, l$ whose mathematical expectation is the true value of
$\sigma^2_{l0}$, but it has a smaller scatter than $\sigma^2_{l0}$.
Let $\bar{\sigma}^2_{l0}$ be this average, we have
then instead of (\ref{Parsevalchi}) a new expression for the reduced
power
\begin{equation}\label{Parsevalchi1}
 \bar{W}_l = W_l\,\frac{\sigma^2_{l0}}{\bar{\sigma}^2_{l0}}
= \frac{{\cal P}_l}{\bar{\sigma}^2_{l0}}\,,
\end{equation}
which is directly proportional to the power and follows a $\chi^2$
distribution with $2l+1$ degrees of freedom.

From a theoretical standpoint, the three indicators are equivalent for
a large number of data points evenly distributed on a sphere. In
practical situations, with some irregularities in the angular
distribution and/or a limited number of sources, the significance
testing with $\bar{W}_l$ should by construction be more robust against
random scatter affecting the estimated standard deviations.
Preliminary experiments have confirmed this feature, although a wider
Monte Carlo simulation is needed to quantify the advantage.

\section{Application to FK5 Catalogue}\label{sect:fk5}

\begin{figure*}[htb]
\begin{center}
\begin{tabular}{cc}
\subfigure[Positions]{
 \includegraphics[width=8cm]{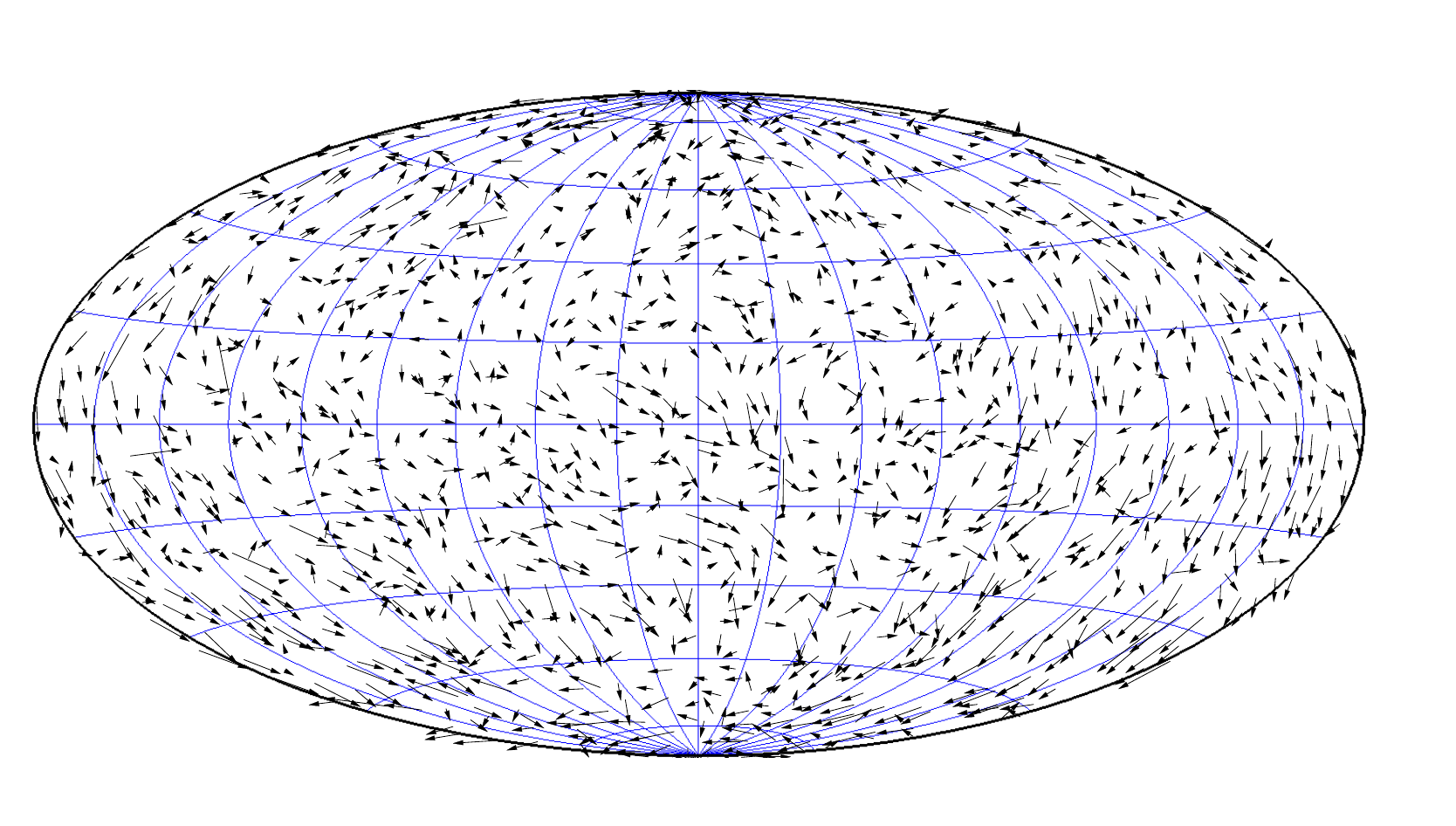}
}
&
\subfigure[Proper motions]{
 \includegraphics[width=8cm]{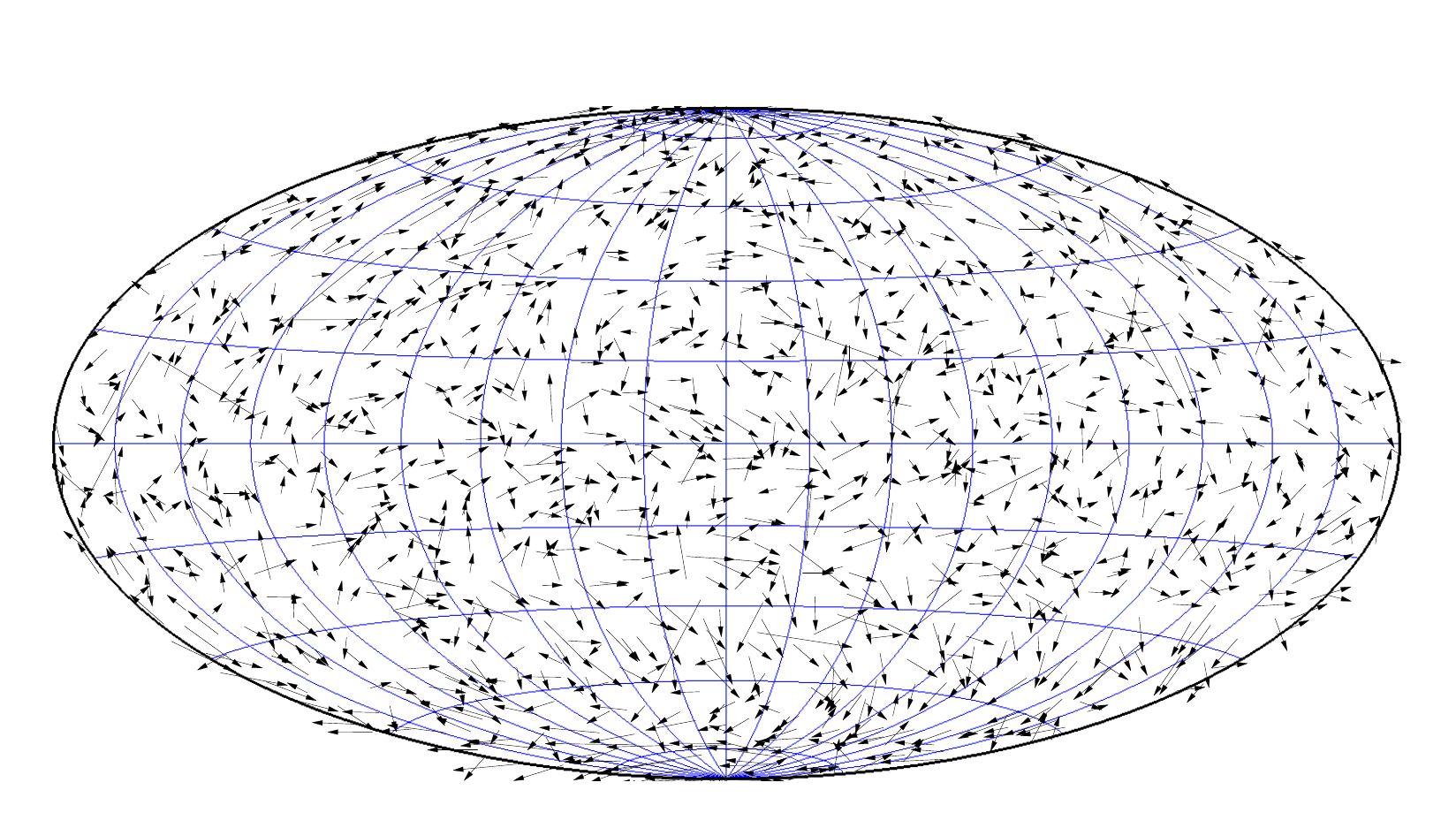}
}
\end{tabular}
\end{center}
 \caption{Differences in positions and proper motions
   between the FK5 catalogue and Hipparcos in 1991.25. The typical
   arrow sizes in the plots are respectively 100 mas and 2.5
   mas/yr.}
\label{fig:FK5-Hipparcos}
   \end{figure*}

The comparison between the FK5 and Hipparcos was published soon after
the release of the Hipparcos catalogue by \citet{Mignard2000b}
primarily provides the orientation and the rotation (spin) of the FK5
system with respect to the Hipparcos frame. The differences in
position and proper motions are shown in Fig.~\ref{fig:FK5-Hipparcos}
for the whole set of stars used in the comparison. In the positional
plot one clearly sees a general shift toward the south celestial pole
and few regional effects of smaller scales.

The global analysis carried out
at that time was restricted to fitting these six parameters from the
positional and proper motion differences from Hipparcos to enable the
transformation of astronomical data from the FK5 frame into the
Hipparcos frame or, more or less equivalently, to the ICRF.  The
remaining residuals were large and considered as zonal errors, not
reducible to a simple representation and presented in the form of
plots of $\Delta \alpha\cos\delta$ and $\Delta\delta$ as a function of
the position of the stars. Large zonal errors were clearly shown
by these plots, but no attempt was made to model these residuals.

Using the VSH now, with the software developed during this
work, it is possible to draw refined conclusions and to show that the
terms of low degree beyond the rotation are truly significant. In
particular we can see in Table \ref{tab:fk5_pos} that the positional
glide term from the spheroidal harmonic of degree $l=1$ is larger than
the orientation term. Terms of degree $l=2$ are also large, both in
position and proper motion, a feature that had not been noticed as
clearly with the earlier analysis based on a priori model.

\subsection{Global differences}

This concerns the rotation and the glide between the two systems, or
equivalently using the VSH terminology, the toroidal and spheroidal
harmonics with
$l =1$. The results for the rotation shown in Table~\ref{tab:fk5_rot}
are fully compatible with the earlier analysis of \citet{Mignard2000b}
and confirm a systematic difference between the FK5
inertial frame and the ICRS materialised by Hipparcos. The
interpretation of the systematic spin is also discussed in the same
paper and seems to be related to the precession constant.  These new
values also agree with a similar analysis performed by
\citet{schwan2001}, provided the epoch is corrected from 1949.4 (mean
epoch of the FK5 observations) to 1991.25 (Hipparcos Catalogue
reference epoch). A new result is given in Table~\ref{tab:fk5_glide}
for the glide, but is only significant in the positions. It describes a
systematic difference of about $60\,\cos\delta$\, mas in the
declination system between the two frames. This is a very large
systematic effect given the claimed accuracy of the FK5 below $50$~mas.
Nothing similar is seen for the proper motions.
Any transformation of a catalogue tied by its construction to
the FK5 system and later on aligned to the HCRF (Hipparcos frame aligned to the
ICRF) must absolutely take this large systematic difference into
account in addition to the rotation. The results just shown correspond
to the harmonics $l=1$ from a fit of the position and proper motion
difference to $l_{\rm max} = 10$. They are slightly different with
another choice of the maximum degree, but within the formal error. In
both tables the standard deviations quoted are derived from the
covariance matrix of the least squares solution scaled by the
r.m.s. of the post-fit residuals.

\begin{table}[hbt]
\setlength{\tabcolsep}{3 mm}
{\small
 \begin{center}
\caption{\em Global orientation (in 1991.25) and spin between the FK5
and the Hipparcos catalogue.}
 \begin{tabular}[h]{ccc@{\qquad}ccc}
\hline
\multicolumn{3}{c} {Position (mas)} & \multicolumn{3}{c}{PM(mas/yr)} \\
     & J1991.25& $\sigma$ & &&  $\sigma$
\\[1 pt]\hline\\[-5pt]
$\epsilon_x$: & -18.1          &  2.4 &  $\omega_x$: & -0.37           & 0.09 \\[1 pt]
$\epsilon_y$: & -14.6          &  2.4 &  $\omega_y$: &\phantom{-}0.57  & 0.09\\[1 pt]
$\epsilon_z$: & \phantom{-}18.5&  2.4 &  $\omega_z$: &\phantom{-}0.82 & 0.09 \\[1 pt]
     \hline \\
     \end{tabular}
   \label{tab:fk5_rot}
 \end{center}
}
\end{table}

\begin{table}[hbt]
\setlength{\tabcolsep}{3 mm}
{\small
 \begin{center}
\caption{\em Glide between the FK5 and the Hipparcos catalogue in
positions (in 1991.25) and proper motions.}
 \begin{tabular}[h]{ccc@{\qquad}ccc}
\hline
\multicolumn{3}{c} {Position (mas)} & \multicolumn{3}{c}{PM(mas/yr)} \\
     & J1991.25& $\sigma$ & & & $\sigma$
\\[1 pt]\hline\\[-5pt]
$g_x$: & \phantom{-}18.3 &  2.4 &  $\gamma_x$: & \phantom{-}0.04  & 0.09 \\[1 pt]
$g_y$: & -1.3            &  2.4 &  $\gamma_y$: & \phantom{-}0.18  & 0.09\\[1 pt]
$g_z$: & -64.0           &  2.4 &  $\gamma_z$: & -0.37  & 0.09 \\[1 pt]
     \hline \\
     \end{tabular}
   \label{tab:fk5_glide}
 \end{center}
}
\end{table}

\subsection{Regional differences}

The analysis at higher degree, up to $l =10$, clearly shows
significant power in most degrees, indicating a definite structure in the
space distribution of the differences. Given the absence of zonal
error in the Hipparcos catalogue, at the level seen in this analysis,
they must come entirely from the FK5.  A detailed analysis of this
regions variations was performed by \citet{schwan2001} by
projecting separately each components (difference in right ascension
and declination) on a set of scalar orthogonal polynomials, instead of
using the vectorial nature of the differences.  Given the numerous
positional instruments used to produce the FK5, some measuring only
the right ascension, others only the declination, others giving both,
analysing the components in equatorial coordinates separately makes
sense and may reveal structure scattered in many degrees with the
VSHs. Here we simply want to illustrate the relevance of the VSH to
analyse the difference between two real catalogues. A detailed
discussion of the application to the FK5 system is deferred to an
independent paper.

\subsubsection{Regional differences in position}

The amplitudes, defined as the square root of the unweighted power, in
each degree for the positional differences between the FK5 and
Hipparcos in 1991.25 are given in Table~\ref{tab:fk5_pos}, for the
toroidal (left part of the table) and spheroidal (right part of the
table) harmonics. The level of significance computed with
Eq.(\ref{Parsevalchi1}) and the normal approximation of
Eq.~(\ref{wilson}) is shown in the columns labelled $Z$. With $l=1$
one recovers the very large significance of the rotation and the
glide. Harmonics $l=2$ are also very large, explaining together with
the glide a significant part of the systematic differences between the
two catalogues. Then the powers remain statistically very significant
up to $l=7$, although with small amplitudes, and then decline. With
the values of the $t_{lm}$ and $s_{lm}$ (not given here), one could
produce an analytical transformation relating the two systems, in the
same spirit as in \citet{schwan2001}. With the last column one sees
that the expansion to $l=10$ explains about $58 \%$ of the variance
seen in the data for the position and $32\%$ for the proper motion.

\begin{table}[hbt]
\setlength{\tabcolsep}{1.2mm}
{\small
 \begin{center}
   \caption{\em Amplitude (power$^{1/2}$) and significance level based
     on (\ref{Parsevalchi1}) for the positional differences between
     FK5 and Hipparcos expanded on the VSH.}
 \begin{tabular}[h]{ccc@{\qquad}ccc}
\hline\\[-8pt]
  degree &  $ \left(\sum_m t^2_{l\,m}\right)^{1/2}$ &  Z & $ \left(\sum_m s^2_{l\,m}\right)^{1/2}$ &  Z   &  Rem. (power)$^{1/2}$ \\
         &           mas                            &    &            mas                          &      &  mas \\
\hline
   0 &            &        &            &        &      474.5\\
   1 &       85.8 &  10.18 &      192.8 &  19.90 &      425.0\\
   2 &       96.6 &  11.48 &      139.5 &  15.94 &      389.6\\
   3 &       53.6 &   6.01 &       52.8 &   5.89 &      382.3\\
   4 &       65.6 &   7.44 &       66.2 &   7.52 &      370.8\\
   5 &       96.8 &  11.41 &       62.2 &   6.72 &      352.5\\
   6 &       65.2 &   6.95 &       82.2 &   9.36 &      336.5\\
   7 &       68.3 &   7.18 &       42.2 &   3.00 &      326.7\\
   8 &       45.0 &   3.18 &       57.1 &   5.22 &      318.6\\
   9 &       31.1 &   0.28 &       41.4 &   2.25 &      314.3\\
  10 &       34.4 &   0.63 &       50.6 &   3.63 &      308.3\\
     \hline \\
     \end{tabular}
\tablefoot{
The last column gives the
power in the remaining signal after the VSH model up to degree
$l$ has been removed. The first line with $l=0$ gives the square
root of the power of the data as defined by
(\ref{power})-(\ref{powernum}).
}
   \label{tab:fk5_pos}
 \end{center}
}
\end{table}

\subsubsection{Regional differences in proper motions}
The same analysis has been done for the proper motions. The
powers and their significance level expressed with a standard normal
variable are listed in Table~\ref{tab:fk5_pm}. Again the signal is
strong in the harmonics of the first two degrees and up to $l=7$
primarily for the toroidal components.

\begin{table}[hbt]
\setlength{\tabcolsep}{1.2mm}
{\small
 \begin{center}
   \caption{\em Amplitude (power$^{1/2}$) and significance level based
     on (\ref{Parsevalchi1}) for the proper motion differences between
     FK5 and Hipparcos expanded on the VSH. }
 \begin{tabular}[h]{ccc@{\qquad}ccc}
\hline\\[-8pt]
  degree &  $ \left(\sum_m t^2_{l\,m}\right)^{1/2}$ &  Z & $ \left(\sum_m s^2_{l\,m}\right)^{1/2}$ &  Z &  Rem. (power)$^{1/2}$\\
         &           mas/yr                         &    &            mas/yr                       &      &  mas/yr   \\
  \hline
     0 &            &        &            &        &       14.6\\
     1 &        3.1 &   9.85 &        1.2 &   3.67 &       14.2\\
     2 &        2.3 &   7.40 &        2.5 &   8.00 &       13.8\\
     3 &        1.1 &   2.10 &        1.5 &   3.88 &       13.7\\
     4 &        1.9 &   5.46 &        1.9 &   5.47 &       13.4\\
     5 &        2.7 &   7.97 &        1.3 &   2.34 &       13.1\\
     6 &        2.0 &   5.16 &        1.8 &   4.33 &       12.8\\
     7 &        2.8 &   8.02 &        1.3 &   1.44 &       12.4\\
     8 &        1.2 &   0.52 &        1.5 &   2.26 &       12.3\\
     9 &        1.3 &   1.17 &        1.2 &   0.28 &       12.2\\
    10 &        1.7 &   2.91 &        1.2 &  -0.04 &       12.0\\
       \hline \\
     \end{tabular}
\tablefoot{
The last column gives the
power in the remaining signal after the VSH model up to degree
$l$ has been removed. The first line with $l=0$ gives the square
root of the power of the data as
defined by (\ref{power})-(\ref{powernum}).
}
   \label{tab:fk5_pm}
 \end{center}
}
\end{table}

\section{Analysis of the expected Gaia results}\label{sect:Gaia-results}

Another important application of the VSH technique is an analysis of
mathematical properties (e.g. systematic errors) of an astrometric
catalogue, as well as extraction of physical information from the
catalogue. Here we concentrate on the QSO catalogue as part of the
expected Gaia catalogue. Our goal is to investigate (i) the
anticipated accuracy of the determination of the rotational state of
the reference frame and the acceleration of the solar system with
respect to the QSOs, and (ii) the expected estimate of the energy flux
of the primordial (ultra-low-frequency) gravity waves.  From the
mathematical point of view, this amounts to checking the accuracy of
determining the VSH coefficients of orders $l=1$ and $l=2$. The
toroidal coefficients of order 1 describe the rotational state of the
reference frame with respect to the QSOs, while the spheroidal
coefficients of order 1 give the acceleration of the solar system with
respect to the QSOs, which shows up as a glide
\citep{Fanselow1983,SoversFanselowJacobs1998,
  kova2003,kopeikin2006,TitovEtAl2011}. The VSH coefficients of order
2 can be related to the energy flux of the ultra-low-frequency gravity
waves \citep{gwinn97}.

\subsection{Simulated QSO catalogue from Gaia}

We simulated the Gaia QSO catalogue with all the details that we know at
the time of writing. The total number of QSOs observable by Gaia --
700\,000 -- was chosen according to the analysis of \citet{Mignard2012}.  The realistic
distribution of the QSOs in the Gaia integrated magnitude $G$ was
taken from \citet{SlezacMignard2007} \citep[see
also][]{RobinEtAl2012}. The distribution of the Gaia-observed QSOs
over the sky was chosen according to the standard Gaia extinction model
\citep{RobinEtAl2012}. The model follows the realistic distribution of
the dust in the Galaxy, so that the QSO distribution used in this
section is not homogeneous on the sky and depends on both angular
coordinates with the most pronounced feature being the avoidance area
close to the Galactic plane. The proper motions of QSOs are generated
from a randomly selected rotation of the reference frame and an acceleration of the
solar system's barycentre with respect to
the QSOs with magnitude and direction expected from the circular
motion of the solar system with respect to the galactic centre
(i.e. the proper motions are generated from the full set of VSH
coefficients of order 1).  These resulted in our simulated
``true'' (noise-free) catalogue of QSOs.

The expected accuracies of astrometric parameters (positions, proper
motions, and parallaxes) as functions of the $G$ magnitude and the
position on the sky are taken from the standard Gaia science
performance model \citep{ESA2011,deBruijne2012}. Then the QSO
simulated solution with Gaia is obtained by adding Gaussian noise with
the corresponding standard deviation to each parameter of each source.

One more feature used in our simulations is that we take
the correlations between the estimates of the two
components of proper motion into account. As a model for the correlations, the
empirical distribution is calculated from the corresponding histogram
for the Hipparcos catalogue as shown on Fig. 3.2.66 of \citet[][vol.1,
Section 3.2]{ESA1997}. We neglect herewith the complicated
distribution of the correlation on the sky as shown on Figs. 3.2.60
and 3.2.61 of \citet{ESA1997}. The generated random correlation is
then used to generate correlated Gaussian noise for the components of
the proper motion of each source.

\subsection{Main results of the analysis of the simulated QSO catalogue}

\begin{figure*}[htb]
\begin{center}
\begin{tabular}{cc}
\subfigure[{\vbox to 5mm{}}Rotation]{
 \includegraphics[width=8cm]{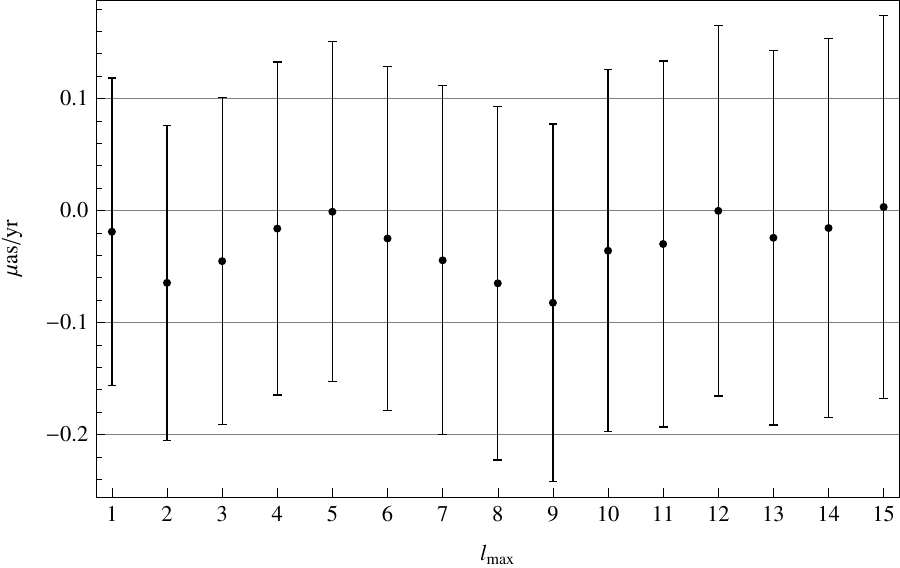}
}
&
\subfigure[{\vbox to 5mm{}}Acceleration (divided by the light velocity)]{
 \includegraphics[width=8cm]{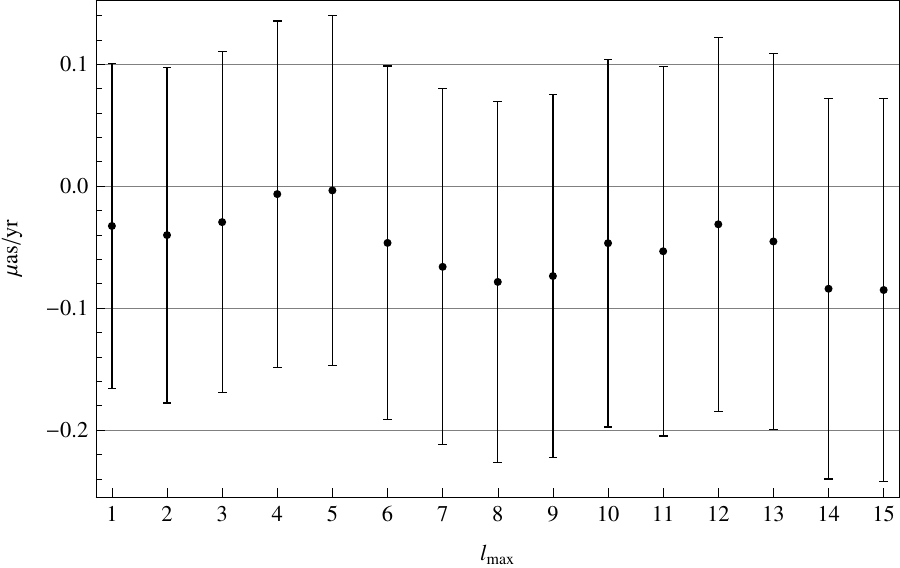}
}
\end{tabular}
\end{center}
\caption{True errors and
  formal errors of the estimates of the absolute values of the
  rotation and acceleration from the fits with $1\le l_{\max}\le15$
  for the simulated QSO catalogue matching expected Gaia accuracy.}
       \label{fig:700kQSOs}
   \end{figure*}

From the resulting realistic QSO catalogue expected from Gaia,
we fitted the VSH coefficients
for maximal order $l_{\max}$  ranging from 1 to 15. The correlations between the
components of the proper motion for the same star are taken into
account in the fit by using block diagonal weight matrix with
$2\times2$ blocks corresponding to each source. Accounting for these
correlations in the fit generally modifies the estimates at the level
of one tenth of the corresponding standard deviations.
We repeated  these simulations (including generation of the QSO catalogue)
tens of times. For a typical simulation,
the resulting biases of the estimates of the rotation of the reference
frame and the acceleration of the solar system, as well as the standard
errors of those estimates are shown in Fig. \ref{fig:700kQSOs}. The
biases are computed as differences between the estimated values and
true ones (those put in the simulated  `noise-free' catalogue). For the standard errors
of the estimates, we took the formal standard deviations from the fit
multiplied by the square root of the reduced $\chi^2$ of the fit.
Comparing of the estimates for different values of
$l_{\max}$ as shown in Fig. \ref{fig:700kQSOs} is a useful indicator
of the reliability of the estimate. Our analysis generally confirms
the assessments of the anticipated Gaia accuracy for the reference
frame and acceleration published previously: both the rotational state
of the reference frame and the acceleration of the solar system could be
assessed with an accuracy of about 0.2 \muas/yr. However, this accuracy
can only be reached under the assumption that (1) Gaia will be
successfully calibrated down to that level of accuracy (that is, the
errors of astrometric parameters remain Gaussian down to about 0.2~\muas),
and (2) real proper motions of the QSOs (transverse
motions of the photocentres due to changing QSO structure, etc.)  are
not too large and do not influence the results. Whether these assumptions
are correct will not be known before Gaia flies.

Finally, we discuss the expected upper estimate of the energy flux
of the ultra-low-frequency gravity waves that Gaia could measure. We use the
computational recipe described by \citet[see Eq. (11) and the
discussion in the first paragraph of their Section 4]{gwinn97}. Our analysis
shows that Gaia can be expected to give an estimate at the level of
$\Omega_{\rm GW}<0.00008\,h^{-2}$ for the integrated flux with
frequencies $\nu<6.4\times 10^{-9}\,{\rm Hz}$. Here, $h=H/(100 {\rm
  km/s/Mpc})$ is the normalised Hubble constant. This is to be
compared with the VLBI estimate $\Omega_{\rm GW}<0.11\,h^{-2}$ for
$\nu<2\times 10^{-9}\,{\rm Hz}$ of \citet{gwinn97}.  The same approach
applied to the VLBI catalogue used by \citet{TitovEtAl2011} gives
$\Omega_{\rm GW}<0.009\,h^{-2}$ for $\nu<1.5\times 10^{-9}\,{\rm
  Hz}$. Therefore, one can expect that Gaia will improve current
estimates from VLBI data by two orders of magnitude while covering a
larger interval of frequencies. However, this conclusion relies on the
ability to see large-scale features
of small amplitude in the Gaia proper motion thanks to the improvement
of the detection in $1/N_\mathrm{sources}^{1/2}$.  This can be
hindered by correlated noise in the intermediates results or
more likely the various sorts of unmodelled systematic errors, not known today,
although every effort is made in the instrument design and
manufacturing to keep them low and compatible with the above
results. These systematic errors can be
expected to influence lower degree VSH coefficients to a greater extent
than the higher degree ones. Therefore, for analysing the gravity waves,
it is advantageous to use not only the
quadrupole harmonics as in \citet{gwinn97} and \citet{TitovEtAl2011},
but also an ensemble of harmonics that includes higher order harmonics
\cite[see e.g.][]{BookFlanagan2011}. Given the number of QSOs in the
Gaia catalogue, such a refined approach is feasible and will help
establish more reliable limits on the energy flux resulting from primordial
gravitational waves.

\subsection{Correlations between the VHS coefficients}

It is interesting to briefly discuss the correlations between the VSH
harmonics resulting from the least squares fits with the simulated catalogue of 700\,000 QSOs
described above. These correlations were computed form the inverse of the weighted normal
matrix from the VHS fit. Depending on the maximal order $l_{\max}$, the root mean square
value of the correlations is 0.04, while a few correlations may attain
the level of 0.4. Having such a large number of sources one might
expect that the correlations should be extremely small.
Indeed, for a catalogue of 700\,000 sources homogeneously distributed on the sky, the r.m.s.
of the correlations between the VSH coefficients is typically 0.0006 and the maximum is at the
level of 0.006. We see that the correlations for the realistic QSO catalogue are
about 70 times larger.  One can find two reasons for the larger correlations:
(1) the inhomogeneous distribution of the QSOs on the sky (this can be
called the ``kinematical inhomogeneity'' of the catalogue) and (2) the fact that the
accuracies of the QSOs' proper motions, which are used in the weight
matrix, also depend on the position on the sky (this can be called the
``dynamical inhomogeneity'' of the catalogue). For the QSO catalogue
used in this section the dynamical inhomogeneity alone gives
correlations with an r.m.s. of 0.01 and the maximum of 0.2, while the
kinematical inhomogeneity alone leads to the r.m.s. of 0.03 and the
maximum of 0.35. Generally speaking, correlations lead to higher
errors in the fitted parameters. We finally note that the
dynamical inhomogeneity will be slightly greater for the real Gaia
catalogue than we see in our simulation. This is caused by the fact
that the standard Gaia science performance model \citep{ESA2011,deBruijne2012}
is smoothed in the galactic longitude. However, our calculations show
that this additional inhomogeneity does not significantly change the
results given above.

\section{Conclusion}

In this paper we have shown the relevance of the vector spherical
harmonics (VSH) to decomposing, more or less blindly, spherical vector
fields frequently encountered in astronomical context.  This happens
very naturally in comparing of stellar positional or proper
motion catalogues or the different solutions produced during the
data analysis in global astrometry. We have provided the mathematical
background and considered several practical issues related to the
actual computation of the VSH.

We have in particular:
\begin{itemize}
  \item shown how to perform the decomposition with a least squares
    minimisation, enabling processing of irregularly distributed sets of
    points on the sphere;
  \item provided a way to test the significance level of any degree in
    the expansion and then set a criterion for where to stop the
    expansion;
  \item stressed the importance of the invariance properties against
    rotation of the expansion and given the necessary formulas
    that express the decompositions in the most common
    astronomical frames;
  \item provided applications with a reanalysis of the FK5 catalogue
    compared to Hipparcos, and an example of how this tool should be
    used during the Gaia data processing to prepare the construction
    of the inertial frame and derive important physical parameters;
  \item compiled an extensive set of practical expressions and
    properties in the Appendices.
\end{itemize}

\begin{acknowledgement}

FM thanks the laboratory Cassiop\'{e}e for supporting his stay at
the Lohrmann Observatory, Dresden, during the early phase of this
work.  SK was partially supported by the BMWi grant 50\,QG\,0901
awarded by the Deutsches Zentrum f\"ur Luft- und Raumfahrt
e.V. (DLR).

\end{acknowledgement}

\appendix

\makeatletter
\if@twocolumn\onecolumn\else\newpage\fi
\makeatother

\section{Explicit formulas for the vector spherical functions with $l\le4$}
\label{annex_1}

\begin{table*}[!hb]
\setlength{\tabcolsep}{2mm}
\begin{center}
\caption{Toroidal harmonics $\vec{T}_{lm}(\alpha, \delta)$
and spheroidal harmonics $\vec{S}_{lm}(\alpha, \delta)$ for $1\le l\le4$.
  \label{tab:Tharm}
  \label{tab:Sharm}
}
{\makeatletter\if@referee\tiny\else\small\fi\makeatother
  \begin{tabular}{cccc}
   \hline \\[-8pt]
   Harm.\,  & Mult.  &   \multicolumn{2}{c}{Components} \\
             & coef.  &   $\ve{e}_\alpha  $  &  $\ve{e}_\delta$ \\[2pt]
   \hline\\[-4pt]
 $\ve{T}_{10}$ & $\frac{1}{2}\sqrt{\frac{3}{2\pi}}$    &   $\cos\delta$                                                & $0$                                                    \\[1mm]
 $\ve{T}_{11}$ & $\frac{1}{4}\sqrt{\frac{3}{\pi}}$     &   $\sin\delta\,(\cos\alpha \,+ \mi\/\sin\alpha)$                          & $-\sin\alpha \,+ \mi\/\cos\alpha$                                 \\[4mm]
 $\ve{T}_{20}$ & $\frac{1}{4}\sqrt{\frac{15}{2\pi}}$   &   $\sin2\delta$                                               & $0$                                                    \\[1mm]
 $\ve{T}_{21}$ & $\frac{1}{4}\sqrt{\frac{5}{\pi}}$     &   $-\cos2\delta\,(\cos\alpha \,+ \mi\/\sin\alpha)$                     & $-\sin\delta\,(\sin\alpha \,- \mi\/\cos\alpha)$                 \\[1mm]
 $\ve{T}_{22}$ & $\frac{1}{8}\sqrt{\frac{5}{\pi}}$     &   $-\sin2\delta\,(\cos2\alpha \,+ \mi\/\sin2\alpha)$                      & $2\cos\delta\,(\sin2\alpha \,- \mi\/\cos2\alpha)$                 \\[4mm]
 $\ve{T}_{30}$ & $\frac{1}{8}\sqrt{\frac{21}{\pi}}$    &   $\cos\delta\,(5\sin^2\delta-1)$                               & $0$                                                    \\[1mm]
 $\ve{T}_{31}$ & $\frac{1}{16}\sqrt{\frac{7}{\pi}}$    &   $\sin\delta\,(15\sin^2\!\delta-11)\,(\cos\alpha \,+ \mi\/\sin\alpha)$   & $-(5\sin^2\!\delta-1)\,(\sin\alpha \,- \mi\/\cos\alpha)$          \\[1mm]
 $\ve{T}_{32}$ & $\frac{1}{8}\sqrt{\frac{35}{2\pi}}$   &   $-\cos\delta\,(3\sin^2\!\delta-1)\,(\cos2\alpha \,+ \mi\/\sin2\alpha)$      & $\sin2\delta\,(\sin2\alpha \,- \mi\/\cos2\alpha)$                \\[1mm]
 $\ve{T}_{33}$ & $\frac{1}{16}\sqrt{\frac{105}{\pi}}$   &   $\cos^2\delta\sin\delta\,(\cos3\alpha \,+ \mi\/\sin3\alpha)$            & $ -\cos^2\!\delta\,(\sin3\alpha \,- \mi\/\cos3\alpha)$                \\[4mm]
 $\ve{T}_{40}$ & $\frac{3}{16}\sqrt{\frac{5}{\pi}}$   &   $\sin2\delta\,(7\sin^2\!\delta-3)$                              & $0$                                                    \\[1mm]
 $\ve{T}_{41}$ & $\frac{3}{16}\sqrt{\frac{1}{\pi}}$    &   $(28\sin^4\!\delta-27\sin^2\!\delta+3)\,(\cos\alpha \,+ \mi\/\sin\alpha)$   & $-\sin\delta\,(7\sin^2\!\delta-3)\,(\sin\alpha \,- \mi\/\cos\alpha)$ \\[1mm]
 $\ve{T}_{42}$ & $\frac{3}{16}\sqrt{\frac{2}{\pi}}$    &   $-\sin2\delta\,(7\sin^2\!\delta-4)\,(\cos2\alpha \,+ \mi\/\sin2\alpha)$     & $\cos\delta\,(7\sin^2\!\delta-1)\,(\sin2\alpha \,- \mi\/\cos2\alpha)$ \\[1mm]
 $\ve{T}_{43}$ & $\frac{3}{16}\sqrt{\frac{7}{\pi}}$    &   $\cos^2\!\delta\,(4\sin^2\!\delta-1)\,(\cos3\alpha \,+ \mi\/\sin3\alpha)$     & $ -3\cos^2\!\delta\sin\delta\,(\sin3\alpha \,- \mi\/\cos3\alpha)$     \\[1mm]
 $\ve{T}_{44}$ & $\frac{3}{8}\sqrt{\frac{7}{2\pi}}$    &   $-\cos^3\!\delta\sin\delta\,(\cos4\alpha \,+ \mi\/\sin4\alpha)$           & $\cos^3\!\delta\,(\sin4\alpha \,- \mi\/\cos4\alpha)$                \\[3mm]
\hline
%
%
   \hline\\[-4pt]
 $\ve{S}_{10}$ & $\frac{1}{2}\sqrt{\frac{3}{2\pi}}$    & $0$                                                    &   $\cos\delta$                                                \\[1mm]
 $\ve{S}_{11}$ & $\frac{1}{4}\sqrt{\frac{3}{\pi}}$     & $\sin\alpha \,- \mi\/\cos\alpha$                                 &   $\sin\delta\,(\cos\alpha \,+ \mi\/\sin\alpha)$                          \\[4mm]
 $\ve{S}_{20}$ & $\frac{1}{4}\sqrt{\frac{15}{2\pi}}$   & $0$                                                    &   $\sin2\delta$                                               \\[1mm]
 $\ve{S}_{21}$ & $\frac{1}{4}\sqrt{\frac{5}{\pi}}$     & $\sin\delta\,(\sin\alpha \,- \mi\/\cos\alpha)$                 &   $-\cos2\delta\,(\cos\alpha \,+ \mi\/\sin\alpha)$                         \\[1mm]
 $\ve{S}_{22}$ & $\frac{1}{8}\sqrt{\frac{5}{\pi}}$     & $-2\cos\delta\,(\sin2\alpha \,- \mi\/\cos2\alpha)$                 &   $-\sin2\delta\,(\cos2\alpha \,+ \mi\/\sin2\alpha)$                        \\[4mm]
 $\ve{S}_{30}$ & $\frac{1}{8}\sqrt{\frac{21}{\pi}}$    & $0$                                                    &   $\cos\delta\,(5\sin^2\!\delta-1)$                               \\[1mm]
 $\ve{S}_{31}$ & $\frac{1}{16}\sqrt{\frac{7}{\pi}}$    & $(5\sin^2\!\delta-1)\,(\sin\alpha \,- \mi\/\cos\alpha)$          &   $\sin\delta\,(15\sin^2\!\delta-11)\,(\cos\alpha \,+ \mi\/\sin\alpha)$       \\[1mm]
 $\ve{S}_{32}$ & $\frac{1}{8}\sqrt{\frac{35}{2\pi}}$   & $- \sin2\delta\,(\sin2\alpha \,- \mi\/\cos2\alpha)$                &   $-\cos\delta\,(3\sin^2\!\delta-1)\,(\cos2\alpha \,+ \mi\/\sin2\alpha)$        \\[1mm]
 $\ve{S}_{33}$ & $\frac{1}{16}\sqrt{\frac{105}{\pi}}$   & $ \cos^2\!\delta\,(\sin3\alpha \,- \mi\/\cos3\alpha)$                &   $\cos^2\!\delta\sin\delta\,(\cos3\alpha \,+ \mi\/\sin3\alpha)$            \\[4mm]
 $\ve{S}_{40}$ & $\frac{3}{16}\sqrt{\frac{5}{\pi}}$   & $0$                                                    &   $\sin2\delta\,(7\sin^2\!\delta-3)$                              \\[1mm]
 $\ve{S}_{41}$ & $\frac{3}{16}\sqrt{\frac{1}{\pi}}$    & $\sin\delta\,(7\sin^2\!\delta-3)\,(\sin\alpha \,- \mi\/\cos\alpha)$   &   $(28\sin^4\!\delta-27\sin^2\!\delta+3)\,(\cos\alpha \,+ \mi\/\sin\alpha)$   \\[1mm]
 $\ve{S}_{42}$ & $\frac{3}{16}\sqrt{\frac{2}{\pi}}$    & $-\cos\delta\,(7\sin^2\!\delta-1)\,(\sin2\alpha \,- \mi\/\cos2\alpha)$ & $-\sin2\delta\,(7\sin^2\!\delta-4)\,(\cos2\alpha \,+ \mi\/\sin2\alpha)$     \\[1mm]
 $\ve{S}_{43}$ & $\frac{3}{16}\sqrt{\frac{7}{\pi}}$    & $3\cos^2\!\delta\,\sin\delta\,(\sin3\alpha \,- \mi\/\cos3\alpha)$     &   $\cos^2\!\delta\,(4\sin^2\!\delta-1)\,(\cos3\alpha \,+ \mi\/\sin3\alpha)$     \\[1mm]
 $\ve{S}_{44}$ & $\frac{3}{8}\sqrt{\frac{7}{2\pi}}$    & $-\cos^3\!\delta\,(\sin4\alpha \,- \mi\/\cos4\alpha)$                &   $-\cos^3\!\delta\,\sin\delta\,(\cos4\alpha \,+ \mi\/\sin4\alpha)$             \\[2mm]
\hline
  \end{tabular}
}
\end{center}
\end{table*}

\makeatletter
\if@referee{}\else\if@onecolumn{}\else\twocolumn\fi\fi
\makeatother

\section{Practical numerical algorithm for the scalar and vector spherical harmonics}
\label{annex_numerical}

\subsection{Scalar spherical harmonics}

Scalar spherical functions $Y_{lm}$ can be computed directly using
definitions (\ref{ylm})--(\ref{ylminusm}).  The Legendre functions can
be evaluated numerically with the following stable recurrence relation
on
\begin{equation}\label{legendre}
    (l-m)\,P_{lm}(x) = (2l-1)\,x\,P_{l-1,m}(x) -(l-1+m)\,P_{l-2,m}
\end{equation}
\noindent
starting with
\begin{eqnarray}\label{startrec}
      P_{m-1,\,m} &=& 0\,,\\
\label{P_mm-explicit}
      P_{mm}   &=& (2m-1)!!(1-x^2)^{m/2}\,.
\end{eqnarray}
\noindent
The algorithm based on (\ref{legendre})--(\ref{P_mm-explicit}) is
discussed in Section 6.8 of \citet{NumRes1992}. The algorithm described there
aims at computing a single value of $P_{lm}(x)$ for given values
of $l\ge0$, $0\le m\le l$ and $|x|\le1$. It is trivial to generalise
it to compute and store all the values of $P_{lm}(x)$ for $l\le l_{\rm max}$,
$l_{\rm max}\ge0$, and a given $x$. The algorithm takes the form
\begin{eqnarray}\label{P_lm-algorithm}
      P_{00} &=& 1,\\
      P_{m+1,m+1} &=& (2m+1)\,\sqrt{1-x^2}\,P_{mm}\,, m=0,\dots,l_{\rm max}-1,\\
      P_{m+1,m} &=& (2m+1)\,x\,P_{mm}\,,m=0,\dots,l_{\rm max}-1,\\
      P_{lm} &=& {1\over l-m}\,\left((2l-1)\,x\,P_{l-1,m}-(l-1+m)\,P_{l-2,m}\right),
\nonumber\\
&&
\qquad
l=m+2,\dots,l_{\rm max},\ m=0,\dots,l_{\rm max}-2.
\end{eqnarray}
\noindent
Here $l_{\rm max}\ge0$ is the maximal value of $l$ to be used in the
computation. The notation ``$a=\dots,\ b=\dots$'' denotes outer cycle for
$b$ and inner cycle for $a$ with the specified boundaries. As a
result one gets a table of values of $P_{lm}(x)$ for a specified $x$ ($|x|\le1$)
and for all $l$ and $m$ such that $0\le m\le l$ and $0\le l\le l_{\rm
  max}$.  An extension of this and related algorithms useful for very
high orders $l$ is given by \citet{Fukushima2012}.

\subsection{Vector spherical harmonics}

To compute vector spherical functions as defined by
(\ref{Tlm})--(\ref{Slm}) one also needs to compute derivatives of the
associated Legendre functions. A closed-form expression for the
derivatives reads as
\begin{equation}\label{legderiv}
   P^\prime_{lm}(x)=\frac{dP_{lm}(x)}{dx} = -\frac{lx}{1-x^2}\,P_{lm}(x) +\frac{l+m}{1-x^2}P_{l-1,m}(x).
\end{equation}
\noindent
Special care should be taken for the computations with $\delta$ close
to $\pm\pi/2$.  Indeed, for $\delta=\pm\pi/2$, one has $x=\sin\delta=\pm1$,
and both the factors $1/\cos\delta$ in (\ref{Tlm})--(\ref{Slm}) and the
derivative $P^\prime_{l1}(x)$ go to
infinity. Therefore, numerical computations (in particular,
Eq. (\ref{legderiv})) become unstable for $\delta$ close to
$\pm\pi/2$.  To avoid this degeneracy the definitions of
$\ve{T}_{lm}$ and $\ve{S}_{lm}$ can be rewritten as
\begin{eqnarray}\label{Tlm-stab}
\ve{T}_{lm}(\alpha,\delta) &=&
(-1)^m \, \sqrt{\frac{2l+1}{4\pi\,l\,(l+1)}\,\frac{(l-m)!}{(l+m)!}}\,
e^{\mi m\alpha}\,
\nonumber\\
&&
\qquad\qquad
\times
\left(A_{lm}(\sin\delta)\,\ve{e}_\alpha - \mi\,B_{lm}(\sin\delta)\,\ve{e}_\delta\right)\,,\\
\label{Slm-stab}
\ve{S}_{lm}(\alpha,\delta) &=&
(-1)^m \, \sqrt{\frac{2l+1}{4\pi\,l\,(l+1)}\,\frac{(l-m)!}{(l+m)!}}\,
e^{\mi m\alpha}\,
\nonumber\\
&&
\qquad\qquad
\times
\left(\mi\,B_{lm}(\sin\delta)\,\ve{e}_\alpha + A_{lm}(\sin\delta)\,\ve{e}_\delta\right)\,,\\
\label{Alm}
A_{lm}(x)&=&\sqrt{1-x^2}\,P^\prime_{lm}(x),\\
\label{Blm}
B_{lm}(x)&=&m\,{1\over\sqrt{1-x^2}}\,P_{lm}(x).
\end{eqnarray}
\noindent
One can easily see that (e.g. from (\ref{Plm})) that $A_{lm}(x)$ and
$B_{lm}(x)$ remain regular for any $0\le m\le l$ and $|x|\le1$. It is easy to see
that numerically stable algorithm for $A_{lm}$ and $B_{lm}$ read as
\begin{eqnarray}
B_{l0}&=&0\,,\ l=1,\dots,l_{\rm max}\,,
\\
B_{11}&=&1\,,
\\
B_{m+1,m+1}&=&{(2m+1)\,(m+1)\over m}\,\sqrt{1-x^2}\,B_{mm},\
\nonumber\\
&&
\qquad\qquad\qquad\qquad\qquad
m=1,\dots,l_{\rm max}-1\,,
\\
B_{m+1,m}&=&(2m+1)\,x\,B_{mm},\ m=1,\dots,l_{\rm max}-1\,,
\\
B_{lm}&=& {1\over l-m}\,\left((2l-1)\,x\,B_{l-1,m} -(l-1+m)\,B_{l-2,m}\right),\
\nonumber\\
&&
\qquad
l=m+2,\dots,l_{\rm max},\ m=1,\dots,l_{\rm max}-2\,.
\\
A_{l0}&=&\sqrt{1-x^2}\,B_{l1},\ l=1,\dots,l_{\rm max}\,,
\\
A_{lm}&=&{1\over m}\,\left(-x\,\,l\,B_{lm}
+(l+m)\,B_{l-1,m}\right),\
\nonumber\\
&&
\qquad\qquad\qquad
m=1,\dots,l,\ l=1,\dots,l_{\rm max}\,.
\end{eqnarray}
\noindent
As a result tables of $A_{lm}(x)$ and $B_{lm}(x)$ are computed for a
given $|x|\le1$, $0\le m\le l$ and $1\le l\le l_{\rm max}$. Explicit
expressions for the vector spherical harmonics for $1\le l\le4$ are
given in Annex \ref{annex_1}.


\section{VSH expansion of a vector field vs. the scalar expansions of
  its components}\label{annex_2}

As explained earlier with the VSH, we can expand a vector field defined
on a sphere on vectorial basis functions preserving the vectorial
nature of the field and behaving very nicely under space
rotations. However, based on the usual practice in geodesy and
gravitation, following \citet{brosche66} and \citet{brosche70}
astronomers have also used, for many years, the expansions of each
component of the vector field, namely $V^\alpha=\ve{V}\cdot\ve{e}_\alpha$ and
$V^\delta=\ve{V}\cdot\ve{e}_\delta$ in the scalar spherical
harmonics. It is interesting to show the relationship between the two
expansions and how their respective coefficients are related to each
other.

\subsection{Definition of the expansions}

Let us consider a vector field $\ve{V}$ on the surface of a sphere.
Its VSH expansion reads as
\begin{equation}\label{Vexpand-VSH}
    \ve{V}(\alpha,\delta) = \sum_{l=1}^\infty\,\sum_{m=-l}^{l}\,
\bigl(t_{lm} \ve{T}_{lm} + s_{lm} \ve{S}_{lm}\bigr)\,.
\end{equation}
\noindent
On the other hand, the components of this vector field
$V^\alpha, V^\delta$  are also expandable independently of each other
in terms of the usual scalar spherical functions $Y_{lm}$:
\begin{eqnarray}\label{Vcomponents1}
V^\alpha&=&\sum_{l=0}^\infty\sum_{m=-l}^l V^\alpha_{lm}\,Y_{lm}\,,
\\\label{Vcomponents2}
V^\delta&=&\sum_{l=0}^\infty\sum_{m=-l}^l V^\delta_{lm}\,Y_{lm}\,,
\end{eqnarray}
\noindent
or for the vector field itself:
\begin{equation}\label{Vexpand-Y}
    \ve{V}(\alpha,\delta) = \sum_{l=0}^\infty\,\sum_{m=-l}^{l}\,
\bigl(
V^\alpha_{lm}\,\ve{e}_\alpha+V^\delta_{lm}\,\ve{e}_\delta
\bigr)\,Y_{lm}\,.
\end{equation}
\noindent
As with all our infinite sums, the
equalities (\ref{Vexpand-VSH})--(\ref{Vexpand-Y}) only hold in the
sense of convergence in the $L^2$ norm as already stated in Section
\ref{subsec:properties}. Pointwise convergence is not guaranteed in
the general case.

Therefore, one has the identity
\begin{eqnarray}\label{VSH-scalar}
\sum_{l=1}^\infty\,\sum_{m=-l}^{l}\,
\bigl(t_{lm} \ve{T}_{lm} + s_{lm} \ve{S}_{lm}\bigr)
=
\sum_{l^\prime=0}^\infty\,\sum_{m^\prime=-l^\prime}^{l^\prime}\,
\bigl(
V^\alpha_{l^\prime m^\prime}\,\ve{e}_\alpha+V^\delta_{l^\prime m^\prime}\,\ve{e}_\delta
\bigr)\,Y_{l^\prime m^\prime}\,.
\end{eqnarray}

\subsection{Formal relations between the coefficients}
Multiplying both sides of (\ref{VSH-scalar}) by $\ve{T}_{lm}^\ast$ or
$\ve{S}_{lm}^\ast$ (again, superscript `$\ast$' denotes complex
conjugation) and integrating over the surface of the sphere, one gets
\begin{eqnarray}\label{tlm-V}
t_{lm} &=& {1\over \sqrt{l\,(l+1)}}\,
\sum_{l^\prime=0}^\infty\,\sum_{m^\prime=-l^\prime}^{l^\prime}\,
\bigl(V^\alpha_{l^\prime m^\prime}\, A_{lml^\prime m^\prime} -
V^\delta_{l^\prime m^\prime}\,B_{lml^\prime m^\prime}\bigr)\,,
\\
\label{slm-V}
s_{lm} &=& {1\over \sqrt{l\,(l+1)}}\,
\sum_{l^\prime=0}^\infty\,\sum_{m^\prime=-l^\prime}^{l^\prime}\,
\bigl(V^\alpha_{l^\prime m^\prime}\, B_{lml^\prime m^\prime} +
V^\delta_{l^\prime m^\prime}\,A_{lml^\prime m^\prime}\bigr)\,,
\end{eqnarray}
\noindent
where
\begin{eqnarray}\label{Almlpmp}
A_{lml^\prime m^\prime} &=&\int_\Omega
{\partial Y^\ast_{lm}\over\partial\delta}\,Y_{l^\prime m^\prime}\,d\Omega,
\\
\label{Blmlpmp}
B_{lml^\prime m^\prime} &=&\int_\Omega
{1\over \cos\delta}\,{\partial Y^\ast_{lm}\over\partial
\alpha}\,Y_{l^\prime m^\prime}\,d\Omega,
\end{eqnarray}
\noindent
where as usual
$d\Omega=\cos\delta\,d\delta\,d\alpha$, and the integration is
taken over the surface of the unit sphere: $0\le\alpha\le2\pi$,
$-\pi/2\le\delta\le\pi/2$.

On the other hand, multiplying both sides of (\ref{VSH-scalar})
by $Y_{lm}^\ast$ and integrating over the surface of the sphere one gets
\begin{eqnarray}\label{Valphalm-ts}
V^\alpha_{lm} &=& \sum_{l^\prime=1}^\infty\,\sum_{m^\prime=-l^\prime}^{l^\prime}\,
{1\over \sqrt{l^\prime(l^\prime+1)}}\,
\bigl(t_{l^\prime m^\prime}\, A^\ast_{l^\prime m^\prime lm} +
s_{l^\prime m^\prime}\,B^\ast_{l^\prime m^\prime lm}\bigr)\,,
\\
\label{Vdeltalm-ts}
V^\delta_{lm} &=& \sum_{l^\prime=1}^\infty\,\sum_{m^\prime=-l^\prime}^{l^\prime}\,
{1\over \sqrt{l^\prime(l^\prime+1)}}\,
\bigl(-t_{l^\prime m^\prime}\, B^\ast_{l^\prime m^\prime lm} +
s_{l^\prime m^\prime}\,A^\ast_{l^\prime m^\prime lm}\bigr)\,.
\end{eqnarray}
\noindent
It remains to compute $A_{l^\prime m^\prime lm}$ and $B_{l^\prime
  m^\prime lm}$ in a convenient way. But formally we have obtained the
two-way correspondence between the coefficients $(t_{lm},s_{lm})$ and
$(V^\alpha_{lm}, V^\delta_{lm})$.

\subsection{Explicit formulas for $A_{lml^\prime m^\prime}$ and $B_{lml^\prime m^\prime}$}

It is straightforward to show that
\begin{eqnarray}\label{Almlpmp-exp}
A_{lml^\prime m^\prime} &=&{1\over 2}\,\pi\,\delta^{mm^\prime}\,\delta^{l+2k+1,l^\prime}\,
\gamma_{lm}\,\gamma_{l+2k+1,m}\,
\nonumber\\
&&
\qquad\qquad\qquad\quad
\times
\left(-l\,\alpha_{lmk}+(l+m)\,
\beta_{l-1,m,k+1}\right)\,,
\\
\label{Blmlpmp-exp}
B_{lml^\prime m^\prime} &=&
-{1\over 2}\,m\,\mi\,\pi\,\delta^{mm^\prime}\,\delta^{l+2k,l^\prime}\,
\gamma_{lm}\,\gamma_{l+2k,m}\,\beta_{lmk}\,,
\\
\label{gammalm}
\gamma_{lm}&=&\sqrt{(2l+1)\,{(l-m)!\over (l+m)!}}\,,
\\
\label{alphalmk}
\alpha_{lmk}&=&{1\over\pi}\,\int_{-1}^{1} {x\over \sqrt{1-x^2}}\,P_{l+2k+1,m}(x)\,P_{lm}(x)\,dx\,,
\\
\label{betalmk}
\beta_{lmk}&=&{1\over\pi}\,\int_{-1}^{1} {1\over \sqrt{1-x^2}}\,P_{l+2k,m}(x)\,P_{lm}(x)\,dx\,,
\end{eqnarray}
\noindent
where $\delta^{ij}$ is the Kronecker symbol ($\delta^{ij}=1$ for $i=j$
and $\delta^{ij}=0$ otherwise) and $k\in\mathbb{Z}$ is arbitrary
integer.
Both $\alpha_{lmk}$ and $\beta_{lmk}$ are positive
rational numbers provided that the indices $l$, $m$, and $k$ are
selected in such a way that the Legendre polynomials under the
integral are different from zero: $|m|\le l$ and $|m|\le l+2k+1$ for
$\alpha_{lmk}$, and $|m|\le l$ and $|m|\le l+2k$ for
$\beta_{lmk}$. Otherwise $\alpha_{lmk}$ and $\beta_{lmk}$ are zero.
Numbers $\alpha_{lmk}$ and $\beta_{lmk}$ can be
computed directly or by using recurrence formulas that can be obtained
from the well-known recurrence formulas for the associated Legendre
polynomials. The following relations allow one to consider only the case
of $\beta_{lmk}$ with positive indices:
\begin{eqnarray}
\alpha_{l,m,-k}&=&\alpha_{l-2k+1,m,k-1}\,,
\\
\beta_{l,m,-k}&=&\beta_{l-2k,m,k}\,,
\\
\alpha_{l,-m,k}&=&{(l-m)!\over(l+m)!}\,{(l-m+2k+1)!\over(l+m+2k+1)!}\,
\alpha_{lmk}\,,
\\
\beta_{l,-m,k}&=&{(l-m)!\over(l+m)!}\,{(l-m+2k)!\over(l+m+2k)!}\,\beta_{lmk}\,,
\\
\alpha_{lmk}&=&
{1\over 2l+4k+3}\,\bigl(\,(l-m+2k+2)\,\beta_{l,m,k+1}
\nonumber\\
&&
\phantom{{1\over 2l+4k+3}\,\bigl(\,+}
+(l+m+2k+1)\,\beta_{lmk}\,\bigr).\,
\end{eqnarray}
\noindent
Finally, a number of equivalent formulas for $\beta_{lmk}$ valid
for any $l\ge0$, $0\le m\le l$ and $k\ge0$ can be derived. Thus, using that
any associated Legendre polynomial can be represented as a finite hypergeometric
polynomial and integrating term by term, one gets the following formula
for $\beta_{lmk}$ valid for any $l\ge0$, $0\le m\le l$ and $k\ge0$:
\begin{eqnarray}
\label{beta-lmk-1}
\beta_{lmk}&=&{(2m-1)!!\over4^m}
\sum_{p=0}^{2(l-m+k)} (-1)^p\,{(2m+2p-1)!!\over 2^p\,(p+2m)!}
\nonumber\\
&&
\qquad\qquad\qquad
\times \sum_{s=\max(0,p-l+m-2k)}^{\min(p,l-m)}
{R_l^{m+s}\,R_{l+2k}^{m+p-s}\over\,s!\,(p-s)!}\,,
\\
R_a^b&=&{(a+b)!\over(a-b)!b!}\,.
\end{eqnarray}
\noindent
We note that a number of explicit formulas for $\beta_{lmk}$ can be
found for $l$, $m$, and $k$ satisfying some specific conditions
(e.g. $m=l$). For the general case, an alternative formula for
$\beta_{lmk}$ can be derived using the Gaunt formula \citep[Appendix,
pp. 192--196]{Gaunt1929}. However, Eq. (\ref{beta-lmk-1}) is
sufficient for practical computations.  These formulas can be easily
implemented numerically.  Equation (\ref{beta-lmk-1}) involves a
double sum of terms of alternating signs.  This means that
computational instabilities can appear if floating point arithmetic is
used.

\subsection{Final relations between the coefficients}

Since $A_{lm l^\prime m^\prime}$ and $B_{lm l^\prime m^\prime}$
vanish unless certain relations between the indices hold, one can
significantly simplify the transformations
(\ref{tlm-V})--(\ref{slm-V}), and
(\ref{Valphalm-ts})--(\ref{Vdeltalm-ts}) can be
significantly simplified and represented as a single sum over
integer $k$:
\begin{eqnarray}
\label{tlm-V-explicit}
t_{lm} &=&
{\pi\,\gamma_{lm}\over 2\sqrt{l\,(l+1)}}\,
\sum_{k\ge {|m|-l-1\over2}}
\bigl(
V^\alpha_{l+2k+1,m}\,p_{lmk}
\nonumber\\
&&
\phantom{{\pi\,\gamma_{lm}\over 2\sqrt{l\,(l+1)}}\,\sum_{k\ge {|m|-l-1\over2}}\bigl(}
+V^\delta_{l+2k,m}\,m\mi\,\gamma_{l+2k,m}\,\beta_{lmk}
\bigr)\,,
\\
\label{slm-V-explicit}
s_{lm} &=&
{\pi\,\gamma_{lm}\over 2\sqrt{l\,(l+1)}}\,
\sum_{k\ge {|m|-l-1\over2}}
\bigl(
-V^\alpha_{l+2k,m}\,m\mi\,\gamma_{l+2k,m}\,\beta_{lmk}
\nonumber\\
&&
\phantom{{\pi\,\gamma_{lm}\over 2\sqrt{l\,(l+1)}}\,\sum_{k\ge {|m|-l-1\over2}}\bigl(}
+V^\delta_{l+2k+1,m}\,p_{lmk}
\bigr)\,,
\\
\label{Valphalm-ts-explicit}
V^\alpha_{lm} &=& {\pi\,\gamma_{lm}\over2}
\sum_{k\ge {\max(|m|,1)-l-1\over 2}}
{1\over \sqrt{l+2k+1}}\,
\Biggl(
t_{l+2k+1,m}\,q_{lmk}
\nonumber\\
&&
\phantom{{\pi\,\gamma_{lm}\over2}
\sum_{k\ge {\max(|m|,1)-l-1\over 2}}}
\qquad
+s_{l+2k,m}\,
{m\mi\,\gamma_{l+2k,m}\over \sqrt{l+2k}}\,\,
\beta_{lmk}
\Biggr)\,,
\\
\label{Vdeltalm-ts-explicit}
V^\delta_{lm} &=&
{\pi\,\gamma_{lm}\over2}
\sum_{k\ge {\max(|m|,1)-l-1\over 2}}
{1\over \sqrt{l+2k+1}}\,
\Biggl(
-t_{l+2k,m}\,
{m\mi\,\gamma_{l+2k,m}\over \sqrt{l+2k}}\,\,
\beta_{lmk}
\nonumber\\
&&
\phantom{{\pi\,\gamma_{lm}\over2}
\sum_{k\ge {\max(|m|,1)-l-1\over 2}}
{1\over \sqrt{l+2k+1}}\,
\Biggl(}
+s_{l+2k+1,m}\,q_{lmk}
\Biggr)\,,
\\
p_{lmk}&=&\gamma_{l+2k+1,m}\,\left(-l\,\alpha_{lmk}+(l+m)\,\beta_{l-1,m,k+1}\right)\,,
\\
q_{lmk}&=&{\gamma_{l+2k+1,m}\over \sqrt{l+2k+2}}\,
\bigl(-(l+2k+1)\,\alpha_{lmk}
\nonumber\\
&&
\phantom{{\gamma_{l+2k+1,m}\over \sqrt{l+2k+2}}\,
\bigl(}
\qquad\qquad
+(l+2k+1+m)\,
\beta_{lmk}\bigr)\,.
\end{eqnarray}
\noindent
In the above formulas we simplified the notations of the lowest values
of $k$ in the sums and assume now that $V^\alpha_{-1,m}=0$,
$V^\delta_{-1,m}=0$, $t_{0m}=0$, and $s_{0m}=0$ and that the terms
containing them vanish identically.

Thus, we have proved that the information of each particular $t_{lm}$ and $s_{lm}$
is distributed over infinite number of $V^\alpha_{l^\prime m}$ and
$V^\delta_{l^\prime m}$ and vice versa.

\subsection{Relation to space rotations}

It is important here to draw attention to two major differences
between the expansions (\ref{Vexpand-VSH}) and
(\ref{Vcomponents1})--(\ref{Vcomponents2}).

\begin{itemize}

\item A rotation between two catalogues is very simply represented in
  the vectorial expansion with the harmonic $\ve{T}_{1m}$ and the
  three coefficients $t_{1,0},\, t_{1,-1},\, t_{1,1}$, while from
  (\ref{Valphalm-ts})-- (\ref{Vdeltalm-ts}), this will require an
  infinite number of coefficients in the component representations. One
  may rightly argue that the opposite is equally true: a simple scalar
  decomposition only over, say, $l=1$ will be much more complex in the
  vectorial expansion. This is true, but physically the really
  meaningful global effects are precisely the rotation and the glide,
  which generate very simple vector fields and project on VSHs of
  first degree only and requires many degrees if done with the scalar
  components. We know of nothing equivalent to the scalar
  representation, unless, obviously, one builds an ad-hoc field  for
  the purpose of illustration.

\item The behaviour of the VSHs or of the scalar spherical harmonics
  $Y_{lm}$ under space rotation are very similar and show the same
  global invariance within a given degree $l$. Their transformations
  are fully defined with the Wigner matrix as shown in
  Section~\ref{wigner}. Therefore it seems at first glance that the
  transformations under space rotation of the $t_{lm}$ and $s_{lm}$ in
  (\ref{Vexpand-VSH}) are not simpler than that of the $V^\alpha_{lm}$
  and $ V^\delta_{lm}$ in
  (\ref{Vcomponents1})--(\ref{Vcomponents2}). There is, however, a very
  important difference that makes the use of the VSH so valuable. By
  applying the Wigner matrix associated to a space rotation to
  (\ref{Vexpand-VSH}), the new coefficients correspond to the expansion
  of the vector field with its components given in the rotated
  frame. Now in (\ref{Valphalm-ts})--(\ref{Vdeltalm-ts}), the two
  scalar fields are considered independently of each other, and the new
  coefficients deduced from the application of the Wigner operator
  correspond to the expansion of the initial scalar fields ($V^\alpha,
  V^\delta$) expressed in the rotated fields, but these components are
  not those of the field $\ve{V}$ projected on the rotated
  coordinates, and they are still the components in the initial frame,
  since a scalar field is invariant by rotation.

\end{itemize}

\end{document}